\documentclass[prr,aps,epsf,twocolumn,longbibliography]{revtex4-1}
\usepackage{times}
\usepackage{graphicx}
\usepackage{float}
\usepackage{latexsym,amsmath,amssymb,bm,euscript}
\usepackage{color}
\usepackage{subfigure}
\usepackage{epstopdf}
\usepackage[colorlinks=true,linkcolor=blue,citecolor=blue,urlcolor=blue]{hyperref}
\usepackage{hyperref}
\usepackage{ulem}
\usepackage[dvipsnames]{xcolor}

\def\ra{\rangle}

\begin{document}

\title{Quantum Transitions of Nematic Phases in a Spin-$1$ Bilinear-Biquadratic Model\\ and Their Implications for FeSe}
\author{Wen-Jun Hu$^{1,2}$}
\author{Hsin-Hua Lai$^{2}$}
\author{Shou-Shu Gong$^{3}$}
\author{Rong Yu$^{4}$}
\author{Elbio Dagotto$^{1,5}$}
\author{Qimiao Si$^{2}$}
\email{qmsi@rice.edu}
\affiliation{
$^1$ Department of Physics and Astronomy, University of Tennessee, Knoxville, Tennessee 37996, USA\\
$^2$ Department of Physics and Astronomy \& Rice Center for Quantum Materials, Rice University, Houston, Texas 77005, USA\\
$^3$ Department of Physics, Beihang University, Beijing 100191, China\\
$^4$ Department of Physics, Renmin University of China, Beijing, 100872, China\\
$^5$ Materials Science and Technology Division, Oak Ridge National Laboratory, Oak Ridge, Tennessee 37831, USA
}

%%%%%%%%
\begin{abstract}
Since its discovery, iron-based superconductivity has been known to develop near an antiferromagnetic order, but this paradigm fails in the iron chalcogenide FeSe, whose single-layer version holds the record for the highest superconducting transition temperature in the iron-based superconductors. The striking puzzle that FeSe displays nematic order (spontaneously broken lattice rotational symmetry) while being non-magnetic, has led to several competing proposals for its origin in terms of either the $3d$-electron's orbital degrees of freedom or spin physics in the form of frustrated magnetism. Here we argue that the phase diagram of FeSe under pressure could be qualitatively described by a quantum spin model with highly frustrated interactions.  We implement both the site-factorized wave-function analysis and the large-scale density matrix renormalization group (DMRG) in cylinders to study the spin-$1$ bilinear-biquadratic model on the square lattice, and identify quantum transitions from the well-known $(\pi,0)$ antiferromagnetic state to an exotic $(\pi,0)$ antiferroquadrupolar order, either directly or through a $(\pi/2,\pi)$ antiferromagnetic state. These many phases, while distinct, are all nematic. We also discuss our theoretical ground-state phase diagram for the understanding of the experimental low-temperature phase diagram obtained by the NMR [P. S. Wang {\it et al.}, Phys. Rev. Lett. 117, 237001 (2016)] and X-ray scattering [K. Kothapalli {\it et al.}, Nature Communications 7, 12728 (2016)] measurements in pressurized FeSe. Our results suggest that superconductivity in a wide range of iron-based materials has a common origin in the antiferromagnetic correlations of strongly correlated electrons.
\end{abstract}

\maketitle

%%%%%%%%%%%
\section{Introduction}

Understanding the iron-based superconductors (FeSCs) has been a subject of extensive research in recent years.~\cite{Kamihara2008, Stewart2011,Si2016,PCDai_review12,Elbio_rmp} The initial interest started with the discovery of superconductivity in the iron pnictides. More recently, iron chalcogenides have provided considerable material variety to this intriguing field and reached the new record of superconducting transition temperature ($T_c$) in FeSCs. These include the potassium iron selenides and other intercalated FeSe systems,~\cite{Guo2010} as well as the single-layer FeSe built on substrates.~\cite{QYWang2012,JJLee_Nature} Because all these record-breaking materials involve FeSe as a building block, it is important to understand the physics of bulk FeSe.~\cite{Wu08,Mao08} Indeed, there is a vast current interest in this system, which possesses the simplest structure among the FeSCs. In contrast to the standard case of the iron pnictides, where a tetragonal-to-orthorhombic structural phase transition is accompanied by a $(\pi,0)$ antiferromagnetic (AFM) order,~\cite{Stewart2011,Si2016,Dai2015} FeSe displays the same type of structural transition, with $T_s \approx 90$ K at ambient pressure, but no magnetic long-range order.~\cite{McQueen2009,Medvedev2009,Bohmer2015,Baek2015,Nakayama2014,Shimojima2014,Watson2015a,Terashima2015} The nematic order is important to a variety of electronic properties of FeSe~\cite{bohmer2017, coldea2018,chen2019,Yi2019}.

Several studies have advanced proposals which attribute the unusual behavior of FeSe to frustrated magnetism among the correlation-induced local moments.~\cite{YuSi_AFQ,FaWang2015,Glasbrenner2015} A non-magnetic, antiferroquadrupolar (AFQ) state with wave vector $(\pi,0)$ appears as a result of frustrated magnetism and has the properties of the bulk FeSe,~\cite{YuSi_AFQ} although additional experimental and theoretical work is needed to confirm the existence of this AFQ state in FeSe. An important appeal of this theoretical picture is that the predicted spin excitations, both for low energies near the wave vector $(\pi,0)$ and for higher energies over an extended range of the Brillouin zone, are compatible with recent experiments.~\cite{Rahn2015,WangZhao2016,QWang2015b, bohmer2017, coldea2018} Meanwhile, parallel proposals~\cite{Bohmer2015,Baek2015,Mukherjee2015} invoke the ordering of the electrons residing on Fe's $3d_{xz}$ and $3d_{yz}$ orbitals, which are degenerate in the $C_4$-symmetric (tetragonal) phase above $T_s$. This idea is also appealing, because the splitting between the two $3d$ orbitals has been observed in angle-resolved photoemission (ARPES) experiments for FeSe.~\cite{Nakayama2014,Shimojima2014,Watson2015a,Yi2019}. Determining which of these competing ideas captures better the essential physics is important to understanding the central microscopic ingredients for the normal state of the FeSCs as well as to elucidate the degree to which the mechanism for superconductivity is universal across the many varieties of FeSCs.

In this paper, we address this issue by exploring the quantum phases and phase transitions related to the nematic phase of FeSe. We focus on studying a spin-$1$ bilinear-biquadratic model on the square lattice, which has been considered before to understand the exotic magnetism and nematic order of the iron-chalcogenide superconductors~\cite{Yu2012,Wysocki2011,FaWang2015,YuSi_AFQ,Wangzhentao2016,gong2017,lai2017,Ruiz2019} but has not been systematically studied to understand the quantum phases and phase transitions of FeSe under pressure. We implement both the site-factorized wave-function analysis and large-scale DMRG method on this model. In general, models with different active microscopic degrees of freedom will have different types of phases in their phase diagrams and, thus, different kinds of quantum phase transitions. In our case, we find four stable spin dipolar and quadrupolar phases, including the N\'{e}el antiferromagnetic order, the $(\pi,0)$ collinear antiferromagnetic phase (CAFM), the $(\pi/2,\pi)$ antiferromagnetic phase (AFM*), and the $(\pi,0)$ antiferroquadrupolar phase (AFQ), and obtain the ground-state phase diagrams. Furthermore, we apply our theoretical results for the understanding of the low-temperature phase diagram that has been indicated by recent experiments of the NMR~\cite{Yuweiqiang2016} and X-ray scattering~\cite{Bohmer2016} measurements in pressurized FeSe. These experiments have demonstrated that lowering temperature induces a tetragonal to orthorhombic (OR) transition, which accompanies a magnetic transition.

It is important to clarify that the actual values of the bilinear and biquadratic couplings, used in our study as free parameters to construct the phase diagrams, could be fixed by analyzing higher level, and far more difficult, multiorbital Hubbard models (see Ref.~\cite{herbrych2018} and references therein) and mapping the low-energy states into the spin-1 model used here. While this study will be carried out in the future, we note that in the bad-metal regime of such a multiorbital setting, the biquadratic interaction is expected to be sizeable~\cite{Book-Fazekas}. Thus, in the present effort generic phase diagrams of the spin-1 bilinear-biquadratic model varying couplings in a broad range will be presented. While future work can clarify with precision where each particular FeSC material is located in our phase diagram, here we focus on whether the overall phase diagram hosts nematic phases and their transitions that pertain to the properties of bulk FeSe.

Our paper is organized as follow. In Sec.~\ref{sec:mm}, we introduce the spin-$1$ bilinear-biquadratic model and describe the computational details. Section~\ref{sec:sfwf} contains the results of the site-factorized wave-function analysis. In Secs.~\ref{sec:numerical} and \ref{sec:nematicity}, we present our DMRG results of the spin and quadrupolar structure factors, as well as the nematic order parameters. Finally, we provide our discussions and conclusions in Sec.~\ref{sec:dis}.

%%%%%%
\section{Model and Methods}\label{sec:mm}

Our starting point is a  spin-$1$ bilinear-biquadratic model on the square lattice,~\cite{Yu2012,Wysocki2011,FaWang2015,YuSi_AFQ,Wangzhentao2016,lai2017} which is defined as
%%%%%%%%%%%%%%%%%%%%%%
\begin{eqnarray}\label{ham}
H= \sum_{i, j} \left[ J_{ij} {\bf S}_i \cdot {\bf S}_j + K_{ij} \left( {\bf S}_i \cdot {\bf S}_j \right)^2 \right] .
\end{eqnarray}
%%%%%%%%%%%%%%%%%%%%%%
Here, ${\bf S}_{i}$ is a spin-$1$ operator at site $i$, which also forms the quadrupolar operator ${\bf Q}_i$, with five independent components:
%%%%%%%%%%%%%%%%%%%%%%
\begin{eqnarray}\label{quad}
&&Q^{x^2-y^2}_i=(S^x_i)^2-(S^y_i)^2, \nonumber\\
&&Q^{3z^2-r^2}_i=[2(S^z_i)^2-(S^x_i)^2-(S^y_i)^2]/\sqrt{3},\nonumber\\
&&Q^{xy}_i=S^x_i S^y_i+S^y_i S^x_i,\nonumber\\
&&Q^{yz}_i=S^y_i S^z_i+S^z_i S^y_i,\nonumber\\
&&Q^{zx}_i=S^z_i S^x_i+S^x_i S^z_i.
\end{eqnarray}
%%%%%%%%%%%%%%%%%%%%%%
The biquadratic term in Eq.~\eqref{ham} can be re-expressed as
%%%%%%%%%%%%%%%%%%%%%%
\begin{eqnarray}
({\bf S}_i \cdot {\bf S}_j )^2 = \frac{1}{2}{\bf Q}_i \cdot {\bf Q}_j-\frac{1}{2}{\bf S}_i\cdot {\bf S}_j+\frac{4}{3}.
\end{eqnarray}
%%%%%%%%%%%%%%%%%%%%%%
In addition, $J_{ij}$ and $K_{ij}$ are respectively the bilinear and biquadratic couplings between the spins at sites $i$ and $j$, with the pair $ij$ denoting distinct bonds on the square lattice. The consideration of biquadratic interaction is quite necessary for spin-$1$ systems. Note that {\it ab initio} method based on density functional theory (DFT) has been used to extract the biquadratic couplings \cite{Glasbrenner2015}.  However, this is a challenging task given that i) FeSe is strongly correlated and ii) quadrupoles, being rank-$2$ objects, do not efficiently couple to the single-particle degrees of freedom \cite{lai2017} that come into the DFT approach. (For related reasons, the DFT simulation may only study different magnetic orders, which cannot explain the non-magnetic phase in the FeSe.) The fact that $J_2$ is comparable with $J_1$ but $K_2$ does not appear in the DFT results illustrates this difficulty~\cite{Glasbrenner2015}. For a spin system with further-neighbor dipolar interactions, it is reasonable to consider the quadrupolar interactions as well.

Following the idea of proposing the $(\pi,0)$ antiferroquadrupolar order phase as the candidate of the non-magnetic phase in the FeSe~\cite{lai2017}, here we consider interactions beyond nearest-neighbor up to the third neighbors. For a minimal model without loss of generality, we consider the nearest-neighbor bilinear interaction $J_1=1$ as the energy unit, with varying second-neighbor interaction $J_2$. In addition, we consider the first three neighbors of the biquadratic interactions to have the same strength for the purpose of simplifying the model and reducing the number of parameters, $-K_1=K_2=-K_3=K>0$. We will demonstrate the robustness of our results by studying the cases with variations of these parameters. For convenience, the present model will be referred to as the $J_1$-$J_2$-$K$ model. In our site-factorized wave-function analysis and DMRG calculations, we can consider both magnetic and non-magnetic phases 
in our model Eq.\eqref{ham}.

For a spin-$1$ model possibly harboring purely magnetic order, purely quadrupolar order, or coexisting magnetic and quadrupolar orders, it is convenient to choose the time-reversal invariant basis of the $SU(3)$ fundamental representation, namely
%%%%%%%%%%%%%%%%%%%%%%
\begin{equation}
\begin{aligned}
|x\ra = \frac{i |1\ra - i |\bar{1}\ra}{\sqrt{2}}, && |y\ra = \frac{ |1\ra + |\bar{1}\ra}{\sqrt{2}}, && |z\ra = -i |0\ra,
\end{aligned}
\end{equation}
%%%%%%%%%%%%%%%%%%%%%%
where we abbreviate $|S^z = \pm1\ra \equiv |\pm1\ra$ $(|S^z = 0\ra \equiv |0\ra)$ and $|\bar{1}\ra \equiv |-1\ra$. Within this basis, the site-factorized wave functions at each site $i$ which characterize any possible ordered state with short-ranged correlations can be expressed as
%%%%%%%%%%%%%%%%%%%%%%
\begin{eqnarray}
|{\bf d}_i\ra = d^x_i |x\ra + d^y_i | y\ra + d^z_i |z\ra,
\end{eqnarray}
%%%%%%%%%%%%%%%%%%%%%%
where $d^x_i$, $d^y_i$, $d^z_i$ are complex numbers and can be re-expressed in the vector form with the basis $\{ |x\ra,~|y\ra,~|z\ra\}$ as $ {\bf d}_i =  ( d^x_i ~ d^y_i ~ d^z_i  )$. It is convenient to separate the real and imaginary parts of ${\bf d}_i$ as ${\bf d}_i = {\bf u}_i + i {\bf v}_i$. The normalization of the wave function leads to the constraint ${\bf d}_i \cdot \bar{\bf d}_i = 1$, or equivalently, ${\bf u}^2_i + {\bf v}^2_i =1$, and the overall phase can be fixed by requiring ${\bf d}_i^2 = \bar{\bf d}_i^2$, i.e., ${\bf u}_i \cdot {\bf v}_i = 0$. In a pure quadrupolar state, ${\bf d}$ will take either a real or imaginary value, but not both, and the associated director is parallel to the director vector ${\bf d}$. This is to be contrasted with a magnetic order, for which ${\bf d}$ contains both real and imaginary components, thus yielding a dipolar magnetic moment. Within this framework, we can determine the spin operator from ${\bf S}_i = 2 {\bf u}_i \times {\bf v}_i$. In terms of the components of ${\bf d}$, the spin and quadrupolar operators can be written as
%%%%%%%%%%%%%%%%%%%%%%
\begin{eqnarray}\label{su3operators}
&&S^\alpha = - i \sum_{\beta \gamma} \epsilon^{\alpha \beta \gamma} \bar{d}^\beta d^\gamma, \nonumber\\
&&Q^{x^2 - y^2} = - |d^x|^2 + |d^y|^2, \nonumber\\
&&Q^{3z^2 - r^2} = \left[ |d^x|^2 + |d^y|^2 -2 |d^z|^2 \right]/\sqrt{3},\nonumber\\
&&Q^{\alpha \beta}|_{\alpha \not=\beta} = - \bar{d}^\alpha d^\beta - \bar{d}^\beta d^\alpha,
\end{eqnarray}
%%%%%%%%%%%%%%%%%%%%%%
with $\alpha/\beta/\gamma=x,y,z$.

%%%%%%%%%%%%%
\begin{figure}[t]
\begin{center}
\includegraphics[width=\columnwidth]{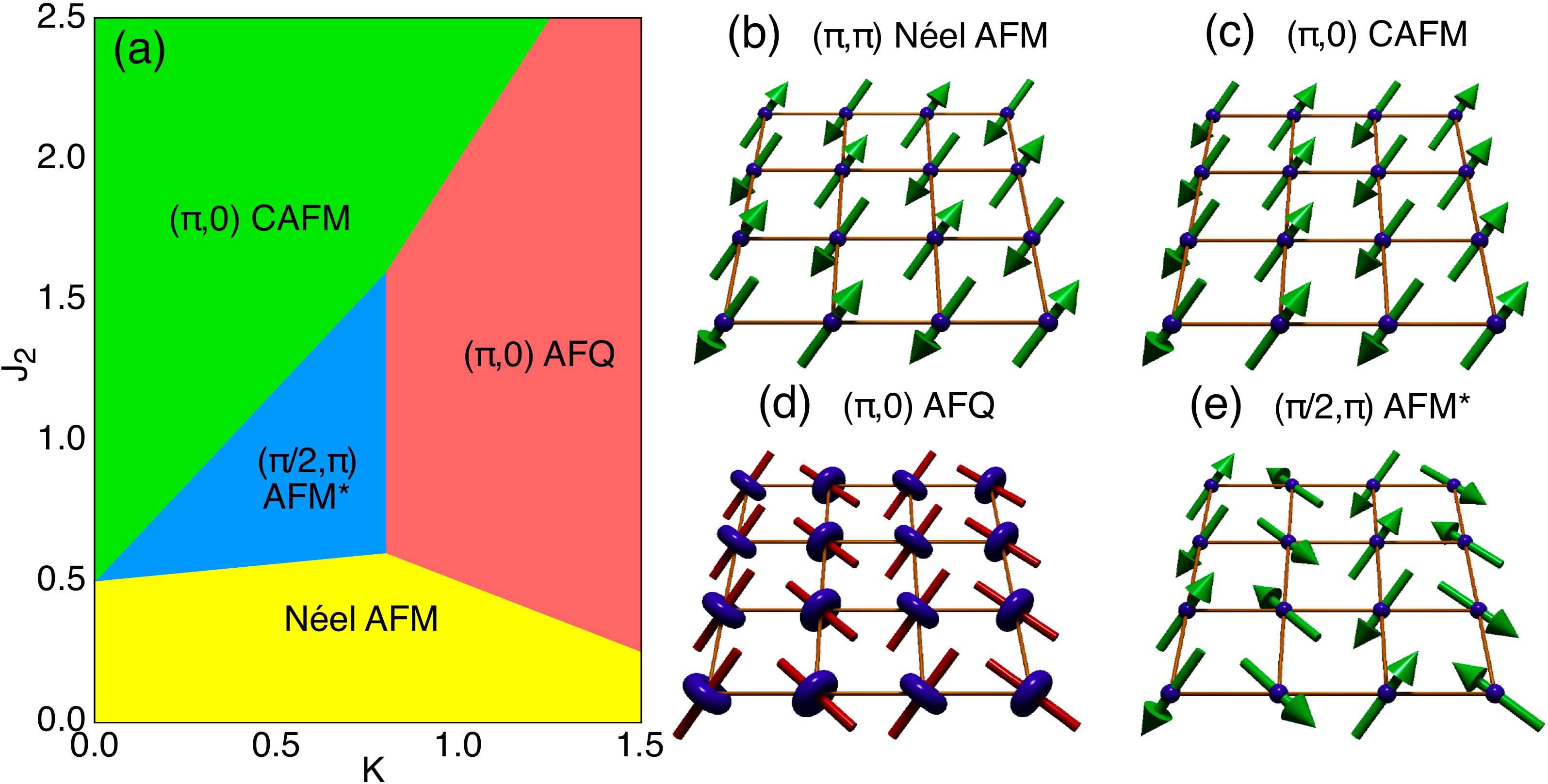}
\end{center}
\caption{(a) Zero-temperature phase diagram of the $J_1$-$J_2$-$K$ model on the $J_2$-$K$ plane ($J_1$ is set to $1$). The phase boundaries are determined from site-factorized wave function calculations. (b)-(e) are schematic illustrations of the four states in (a), including N\'{e}el AFM (b), $(\pi,0)$ CAFM (c), $(\pi,0)$ AFQ (d), and $(\pi/2,\pi)$ AFM$^*$ (e). The green arrows denote the spins. The red thin cylinders in (e) describe the quadrupolar directors and the blue donut-shaped objects represent the spin fluctuations which are perpendicular to the directors.}
\label{mfpd}
\end{figure}
%%%%%%%%%%%%%

In addition to the site-factorized wave-function analysis, we also study the model Eq.~\eqref{ham} by the density matrix renormalization group (DMRG) with spin rotational $SU(2)$ symmetry. We perform the DMRG simulations on $L \times 2L$ cylindrical systems with $L = 4,6,8$ in the $y$ direction. The cylinder geometry used here has open boundary conditions along the $x$ direction and periodic boundary conditions along the $y$ direction. We keep up to $4000$ $SU(2)$ DMRG states. In the N\'{e}el AFM and the $(\pi,0)$ collinear antiferromagnetic phase (CAFM), the truncation error is around $10^{-6}$, while in the $(\pi/2,\pi)$ AFM$^*$ and the $(\pi,0)$ antiferroquadrupolar phase (AFQ) the truncation error is around $10^{-5}$. These small truncation errors ensure us to obtain accurate DMRG results.

%%%%%%%%%%%%%%%
\section{Site-factorized wave-function analysis}\label{sec:sfwf}

To explore the possible quantum phases of the model Eq.~\eqref{ham}, we start from an analysis based on a site-factorized wave-function analysis.~\cite{Bauer2012,YuSi_AFQ,Wangzhentao2016,lai2017} In this framework, we can re-express the model Hamiltonian as
%%%%%%%%%%%%%
\begin{eqnarray}\label{Eq:H_d}
H = \sum_{i,j} \left[J_{ij}\left| {\bf d}_i \cdot \bar{\bf d}_j\right|^2
+  \left(K_{ij} - J_{ij}\right) \left|{\bf d}_i \cdot {\bf d}_j \right|^2 + K_{ij}\right].
\end{eqnarray}
%%%%%%%%%%%%%
In order to obtain the variational phase diagram, we should numerically minimize the Hamiltonian above. In the present analysis, we consider four stable spin dipolar or quadrupolar ordered phases as illustrated in Fig.~\ref{mfpd}, including the CAFM, N\'{e}el AFM, and $(\pi,0)$ AFQ, as well as a newly discovered magnetic phase dubbed $(\pi/2,\pi)$ AFM$^*$. Note that here we do not consider the ferroquadrupolar order discussed in Ref.~\cite{Wangzhentao2016} as a candidate for the non-magnetic phase but only consider the $(\pi,0)$ AFQ order. From the perspective of purely theoretical explorations, the two types of quadrupolar states are both intriguing. However, the ferroquadrupolar order itself does not generate a nematic order.

The N\'{e}el AFM in Fig.~\ref{mfpd}(b) shows the conventional spin pattern with the nearest-neighbor spins anti-parallel to each other. In the CAFM phase, shown in Fig.~\ref{mfpd}(c), the nearest-neighbor spins are parallel to each other along one direction, while they are antiparallel to each other along the other direction. The $(\pi,0)$ AFQ with staggered quadrupolar order along one direction, as shown in Fig.~\ref{mfpd}(d), can be characterized by having mutually orthogonal nearest-neighbor directors, i.e., ${\bf d}_i \cdot {\bf d}_j = 0 = {\bf d}_i \cdot \bar{\bf d}_j$.

The novel $(\pi/2, \pi)$ AFM$^*$, shown in Fig.~\ref{mfpd}(e), is a new phase where the spin direction along the $x$-axis rotates with a commensurate period of four sites while its period along the $y$-axis is still two sites reflecting the wave vector $(\pi/2,\pi)$. This AFM$^*$ state is nematic since it spontaneously breaks the lattice $C_4$ symmetry, by choosing between two degenerate wave vectors $\mathbf{q} = (\pi/2,\pi)$ and $(\pi, \pi/2)$. As an illustration, the spin configuration of the $(\pi/2, \pi)$ AFM$^*$ state along the $x$-axis may take the $4$-site periodic pattern as $\{ |S^z = 1\ra$, $|S^x = 1\ra$, $|S^z = -1\ra$, $|S^x = -1\ra\}$, while the spin orientation still takes the conventional staggered pattern along the $y$-axis. Importantly, in the $(\pi/2, \pi)$ AFM$^*$ phase, the ${\bf q}=(\pi,0)$ quadrupolar order parameter is nonzero as well, which is responsible for the stability of this phase.

Within the site-factorized wave-function studies, the energy per site of each phase can be obtained (see Appendix~\ref{app2}) as follows
\begin{eqnarray}
&& \mathcal{E}^{N\acute{e}el}_{AFM} = 2 \left(K_1 - J_1\right) + 2 J_2 + 2 J_3=-2 + 2J_2 - 2K,\nonumber\\
&& \mathcal{E}_{CAFM} = K_1 + 2 (K_2 - J_2) + 2J_3 = -2J_2 + K,\nonumber\\
&& \mathcal{E}_{(\pi/2,\pi)AFM^*} = -J_1 + \frac{5}{4}K_1 + \frac{1}{2}K_2 + K_3=-1 - \frac{7}{4} K,\nonumber\\
&& \mathcal{E}_{(\pi,0)~AFQ} = K_1 + 2K_3=-3K.\nonumber
\end{eqnarray}
For all energies, we have neglected the constant term $K_{ij}$ in Eq.~\eqref{Eq:H_d}. Note that the SU(3) analysis was considered earlier, in particular in Ref.\,\onlinecite{lai2017}. It is also worth emphasizing that these results capture the quantum fluctuations inherent to the $S=1$ case, as discussed in some detail in Appendix~\ref{app2}. Using these energies, we can analytically determine the boundaries as follows
\begin{enumerate}
\item[(1)]Phase boundary between N\'{e}el AFM and $(\pi,0)$ AFQ: $K + 2J_2 - 2 = 0.$
\item[(2)]Phase boundary between N\'{e}el AFM and $(\pi/2,\pi)$ AFM$^*$: $K - 8J_2 + 4 = 0.$
\item[(3)]Phase boundary between $(\pi/2, \pi)$ AFM$^*$ and CAFM: $11K - 8J_2 + 4 = 0.$
\item[(4)]Phase boundary between $(\pi/2, \pi)$ AFM$^*$ and $(\pi,0)$ AFQ: $5K - 4 = 0.$
\item[(5)]Phase boundary between CAFM and $(\pi,0)$ AFQ: $2K - J_2 = 0.$
\end{enumerate}
Using these equations, by employing the site-factorized wave-function analysis we obtain the variational phase diagram shown in Fig.~\ref{mfpd}(a).

We would like to mention that within the site-factorized wave-function analysis, a phase with coexistent magnetic and quadrupolar orders at different real-space sites, dubbed AFMQ, has been found by minimizing the energy. To our best understanding, the AFMQ shows the same period as that of $(\pi/2, \pi)$ AFM$^*$, but is an inhomogeneous phase with finite magnetic and quadrupolar orders at different real-space columns (rows). The spin pattern in AFMQ presents a staggered pattern between magnetically-ordered sites, while the quadrupolar pattern is ferroquadrupolar (FQ), i.e. the quadrupolar directors are all parallel to each other. The site-factorized wave-functions between the magnetically-ordered sites and the quadrupolar sites are orthogonal to each other. Since this regime with coexisting magnetic and quadrupolar orders at different columns (rows) appears between the $(\pi/2,\pi)$ AFM$^*$ and $(\pi,0)$ AFQ, whose period is consistent with both $(\pi/2, \pi)$ AFM$^*$ and $(\pi,0)$ AFQ, it is likely that this phase is just a transition regime between the purely magnetic phase and the purely quadrupolar phase. Its existence reflects the first-order nature of the transition between the two phases, and it is expected to be destabilized by quantum fluctuations; this is confirmed by our DMRG calculations later. We have therefore ignored this regime in the phase diagram displayed in Fig.~\ref{mfpd}(a).

%%%%%%%%%%%%%
\begin{figure}[t]
\begin{center}
\includegraphics[width=\columnwidth]{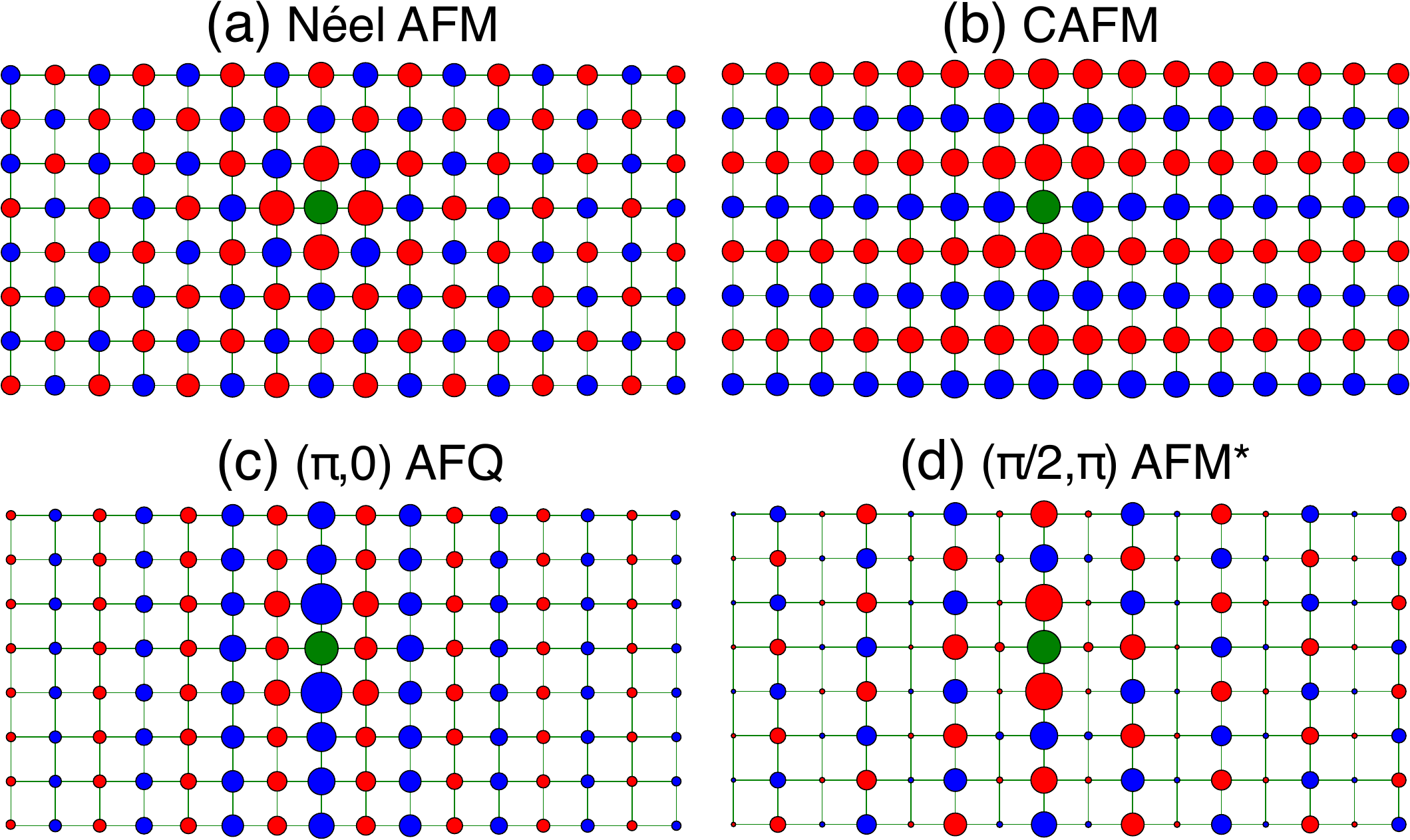}
\end{center}
\caption{(a) The spin-spin correlations for the N\'{e}el AFM state at $J_2=0.4$ and $K=0.6$. (b) The spin-spin correlation for the CAFM state at $J_2=1.5$ and $K=0.5$.  (c) The quadrupolar-quadrupolar correlation for the $(\pi,0)$ AFQ state at $J_2=1.5$ and $K=0.7$. (d) The spin-spin correlation for the ($\pi/2,\pi$) AFM$^*$ state at $J_2=0.8$ and $K=0.35$. The green site is the reference site; the blue and red colors denote positive and negative correlations of the sites with the reference site, respectively. The area of circles is proportional to the magnitude of the spin or quadrupolar correlation.}
\label{realspace}
\end{figure}
%%%%%%%%%%%%%

%%%%%%%%%%%%%
\begin{figure}[t]
\begin{center}
\includegraphics[width=\columnwidth]{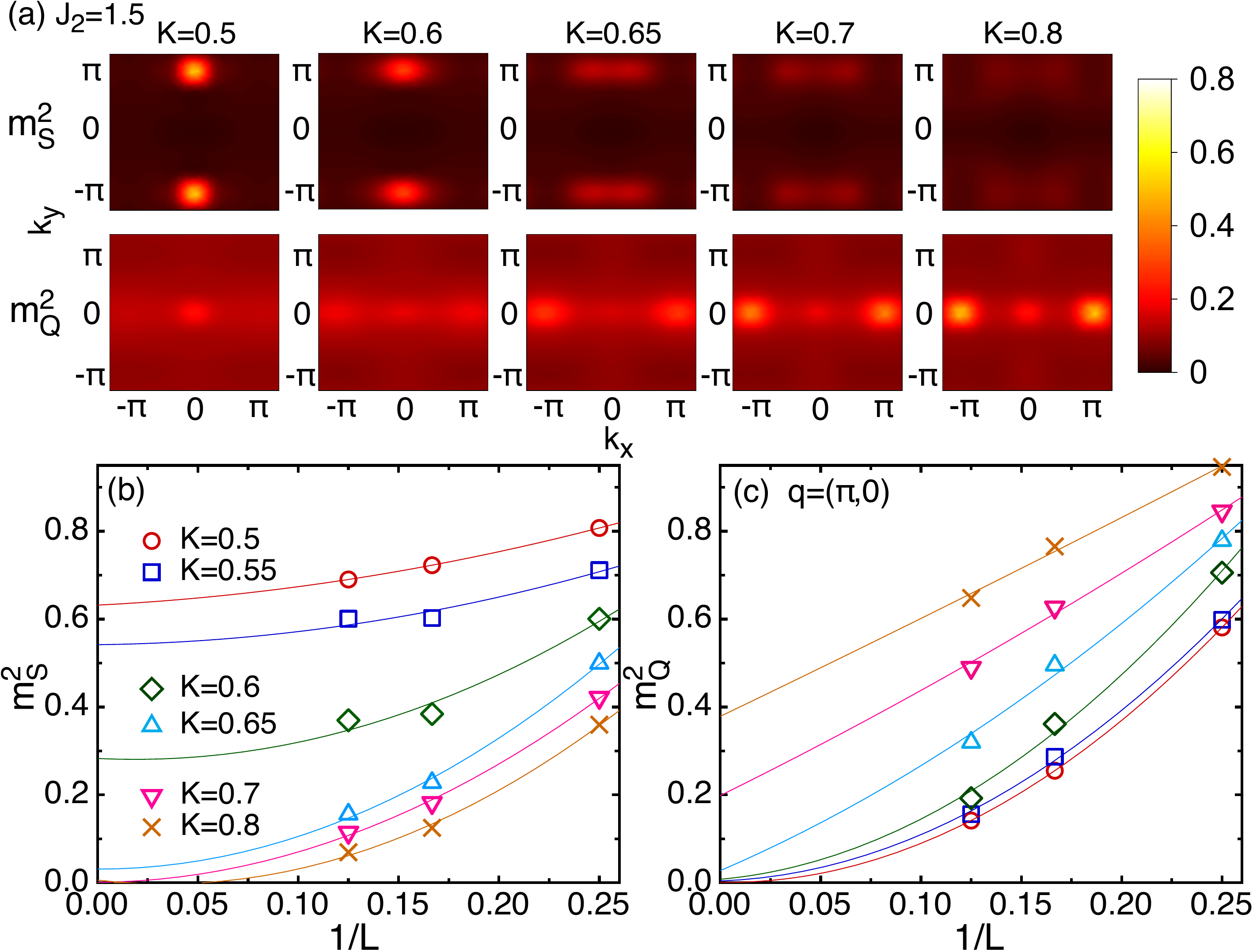}
\end{center}
\caption{(a) Spin ($m^2_{S}$) and quadrupolar ($m^2_{Q}$) structure factors obtained from the DMRG calculations on the $8\times 16$ cylinders for $J_2=1.5$. Both structure factors display dramatic changes at $K \simeq 0.65$, indicating a phase transition from the $(0,\pi)$ CAFM to the $(\pi,0)$ AFQ. In the $(\pi,0)$ AFQ, $m^2_Q$ exhibits a characteristic peak at $(\pi,0)$. Finite-size scaling for the spin (b) and quadrupolar (c) structure factors at different values of $K$ for $J_2 = 1.5$. For the spin structure factor (b), the highest peak of $m^{2}_S$ in its momentum distribution is shown. For the quadrupolar structure factors (c), the intensity at ${\bf q}=(\pi,0)$ is plotted. According to the scaling, the value $K=0.65$ is close to the phase boundary. Lines are guides to the eye.}
\label{dmrg15}
\end{figure}
%%%%%%%%%%%%%

%%%%%%%%%%%%%
\begin{figure}[t]
\begin{center}
\includegraphics[width=\columnwidth]{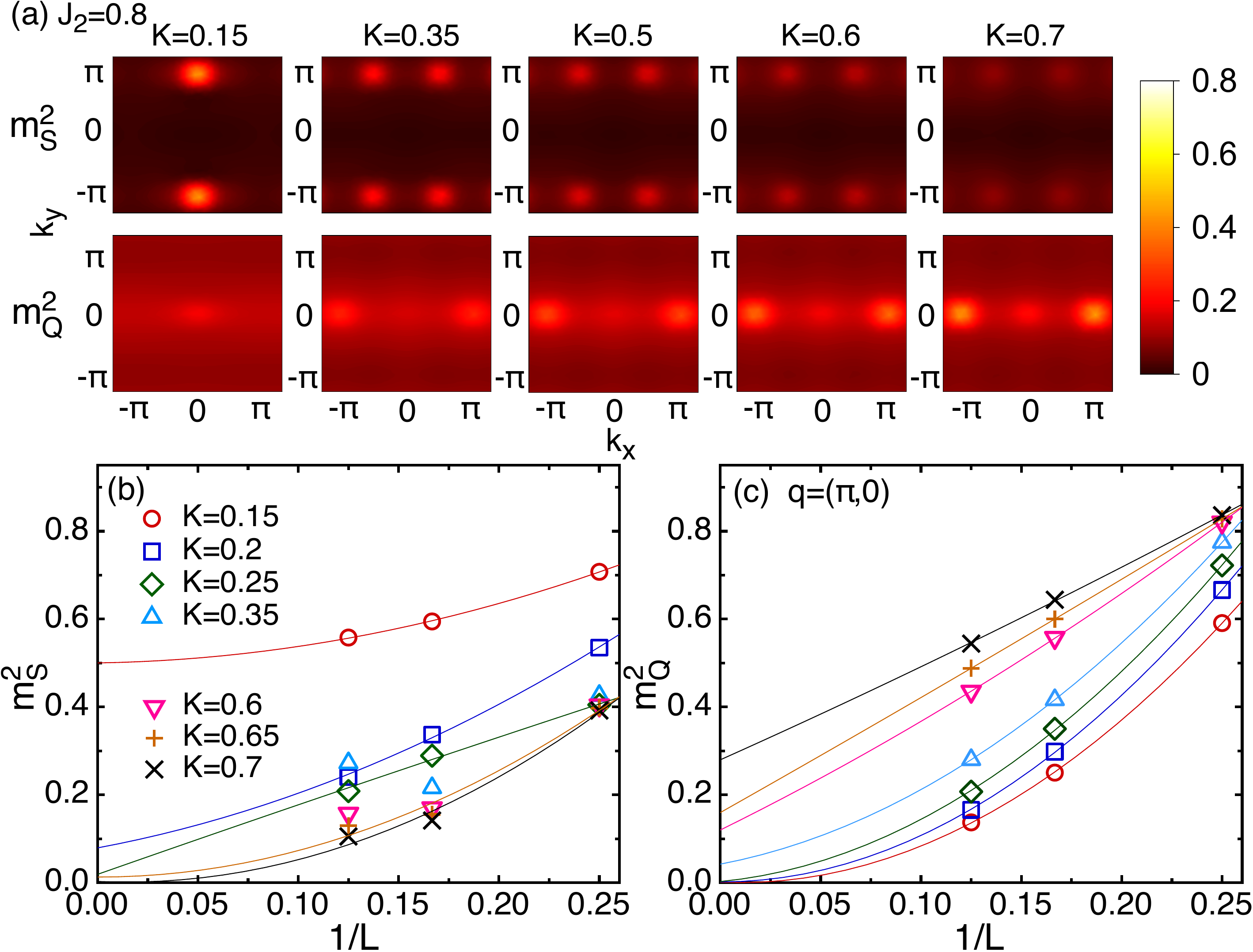}
\end{center}
\caption{(a) Spin ($m^2_{S}$) and quadrupolar ($m^2_{Q}$) structure factors obtained from the DMRG calculations on the $8\times 16$ cylinders for $J_2=0.8$. At this value of $J_2$, $m^2_S$ displays a transition from $(0,\pi)$ CAFM, through $(\pi/2,\pi)$ AFM$^*$, and finally to $(\pi,0)$ AFQ. In both $(\pi/2,\pi)$ AFM$^*$ and $(\pi,0)$ AFQ, $m^2_Q$ exhibits a characteristic peak at $(\pi,0)$. Finite-size scalings for the spin (b) and quadrupolar (c) structure factors at different values of $K$ and at fixed $J_2 = 0.8$. For the spin structure factor (b), the highest peak of $m^{2}_S$ in its momentum distribution is shown. For the quadrupolar structure factors (c), the intensity at ${\bf q}=(\pi,0)$ is plotted. The momentum $(\pi/2,\pi)$ is not an allowed lattice vector on the $6\times 6$ cluster, which is responsible for the apparent non-monotonic dependence of $m^2_S$ vs $1/L$ in the range $0.25<K<0.65$. Lines are guides to the eye.}
\label{dmrg08}
\end{figure}
%%%%%%%%%%%%%

%%%%%%%%%%%%%%%%%%%%%%%%%%%%
\section{DMRG Phase Diagrams}\label{sec:numerical}

Our analysis so far has been semi-classical. In order to explore the role of {\it full} quantum fluctuations and analyze the model of Eq.~\eqref{ham} in an unbiased way, we have also carried out large-scale density matrix renormalization group (DMRG) calculations.~\cite{White1992} First of all, we selected four points of the four stable phases shown in the phase diagram Fig.~\ref{mfpd}(a): $J_2=0.4$ and $K=0.6$ for the N\'{e}el AFM, $J_2=1.5$ and $K=0.5$ for the CAFM, $J_2=1.5$ and $K=0.7$ for the $(\pi,0)$ AFQ, and $J_2=0.8$ and $K=0.35$ for the ($\pi/2,\pi$) AFM$^*$. Next, we computed the spin-spin ($\langle {\bf S}_{i}\cdot {\bf S}_{j} \rangle$) and quadrupolar-quadrupolar ($\langle {\bf Q}_{i}\cdot {\bf Q}_{j} \rangle$) correlation functions for these four points by using DMRG on a $L=8$ cylindrical geometry, which contain the real-space spin and quadrupolar configurations displayed in Fig.~\ref{realspace}. The spin pattern of the N\'{e}el AFM state is in Fig.~\ref{realspace}(a); the CAFM state in Fig.~\ref{realspace}(b) automatically chooses the anti-parallel configuration along the $y$-direction and the parallel configuration along the $x$-direction, due to the cylindrical geometry; the AFQ phase in Fig.~\ref{realspace}(c) has the antiferroquadrupolar configuration along the $x$-direction and the ferroquadrupolar configuration along the $y$-direction. In the $(\pi/2,\pi)$ AFM$^*$ phase, along the $y$-direction the spin configuration is anti-parallel, while  along the $x$-direction the spin pattern in Fig.~\ref{realspace}(d) indicates that the spins are orthogonal between two nearest-neighbor sites. This selection of otherwise degenerate nematic states is induced by the small symmetry-breaking geometry of the cylinders used in DMRG.

Next we consider the evolutions of the zero-temperature phases as a function of the biquadratic $K$ coupling to obtain the ground-state phase diagram for fixed $J_2$. In order to identify the phases we encounter, we have calculated the static spin and quadrupolar structure factors defined as
%%%%%%%%%%%%%
\begin{eqnarray}\label{m2}
&&m^{2}_{S}({\bf q})=\frac{1}{L^4}\sum_{i,j} \langle {\bf S}_{i}\cdot {\bf S}_{j} \rangle e^{i{\bf q}\cdot({\bf r}_i-{\bf r}_j)}, \\
&&m^{2}_{Q}({\bf q})=\frac{1}{L^4}\sum_{i,j} \langle {\bf Q}_{i}\cdot {\bf Q}_{j} \rangle e^{i{\bf q}\cdot({\bf r}_i - {\bf r}_j)},
\end{eqnarray}
%%%%%%%%%%%%%
where $i,j$ are restricted to be only partially summed over the $L \times L$ sites in the {\it middle} of the cylinder so that the finite-size effects are reduced.~\cite{gong2014square}

Figure~\ref{dmrg15} displays the DMRG results for $J_2=1.5$. Following the scaling behavior in magnetic order states which can be obtained from spin-wave theory, we show the order parameters as a function of $1 / L$. Note that, given the numerically intensive nature of the DMRG  calculations for an extended parameter space, we only crudely extrapolate the data for the finite-size scaling, which is sufficient to show whether the order vanishes or not. The evolutions of the spin ($m^2_S$) and quadrupolar ($m^2_Q$) structure factors is shown in Fig.~\ref{dmrg15}(a). Here, in the range $K<0.65$ the peak at momentum ${\bf q}=(0,\pi)$ of the spin structure factor suggests the presence of the CAFM phase while the quadrupolar structure factor has its peak at ${\bf q}=(0,0)$. On the other hand, for $K>0.65$ the quadrupolar structure factor develops a clear peak at ${\bf q}=(\pi,0)$, while the peak in the spin structure factor has melted, indicating the $(\pi,0)$ AFQ order. Due to the cylindrical geometry, the CAFM state automatically selects the configuration with ${\bf q}=(0,\pi)$, whereas the AFQ phase selects ${\bf q}=(\pi,0)$. Furthermore, we examine the finite-size scalings of the spin ($m^{2}_S$) and quadrupolar ($m^{2}_Q$) order parameters at different values of $K$ in Figs.~\ref{dmrg15}(b) and (c). Upon increasing $K$, the spin order parameter $m^{2}_S(0,\pi)$ decreases and vanishes when $K>0.65$. Instead, the quadrupolar order parameter $m^{2}_Q(\pi,0)$ develops for $K>0.65$. The behavior of both these order parameters indicate a direct phase transition from CAFM to $(\pi,0)$ AFQ, and $K=0.65$ is roughly the location of the phase boundary based on the finite-size scalings. Beyond the discovery of the $(\pi,0)$ AFQ phase as a genuine ground state in~\cite{lai2017}, here we show that the CAFM and $(\pi,0)$ AFQ appear as nearby phases through the variation of the biquadratic couplings.

We also performed DMRG calculations at fixed $J_2 = 0.8$, with results in Fig.~\ref{dmrg08}. At small and large values of $K$, the system is in the CAFM and ($\pi,0$) AFQ phases, respectively. In the intermediate region, there is a new magnetic phase emerging, with a new peak in the spin structure factor $m^2_S$ developing at ${\bf q} = (\pi/2,\pi)$ as shown in Fig.~\ref{dmrg08}(a), for example at $K=0.35$. Also, we find the peak at ${\bf q} = (\pi,0)$ for the quadrupolar structure factor $m^2_Q$. These results are consistent with our site-factorized wave-function analysis for the $(\pi/2,\pi)$ AFM$^*$ order in Section~\ref{sec:sfwf}, which suggests the coexistence of the $(\pi/2,\pi)$ magnetic and $(\pi,0)$ quadrupolar orders.  Figs.~\ref{dmrg08}(b) and (c) contain the finite-size scalings of spin $m^{2}_{S}$ and quadrupolar $m^{2}_{Q}$ order parameters at different values of $K$. Although the momentum $(\pi/2,\pi)$ is not an allowed lattice vector on the $6\times 6$ cluster, and creates the apparently non-monotonic dependence of $m^2_S$ vs $1/L$ in the range $0.25<K<0.65$ shown in Fig.~\ref{dmrg08}(b), still the results clearly demonstrates the nonzero values of both order parameters $m^{2}_{S}$ and $m^{2}_{Q}$ at $K \sim 0.6$ after the finite-size scaling. This signature is less clear for smaller $K$, but we believe that both orders already coexist there. As inferred from the finite-size scalings, we find two quantum phase transitions, with the first one happening around $K \simeq 0.25$ and the second around $K \simeq 0.65$.

We would like to mention that the particular parameter cut in the model is for presentation purposes only. The nematic phases actually span over large parameter regions, and the general features of the nematic phase diagram remain the same if we choose other fixed parameters (see Appendix~\ref{app1}).

%%%%%%%%%%%%%
\begin{figure}[t]
\begin{center}
\includegraphics[width=\columnwidth]{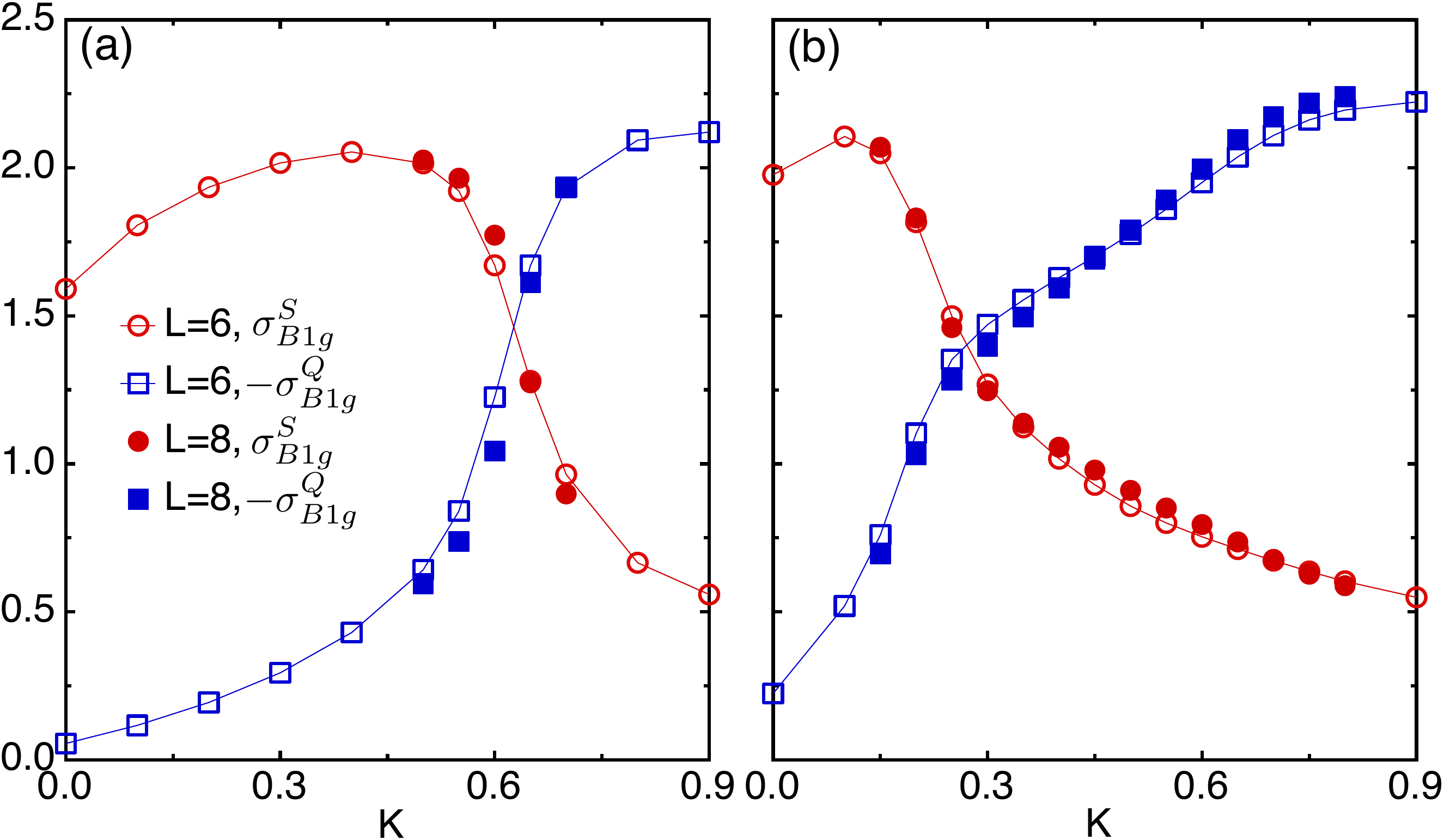}
\end{center}
\caption{The spin and quadrupolar nematic order parameters [$\sigma^S_{B1g}$ and $\sigma^Q_{B1g}$, defined in Eqs.~(\ref{sigma1}) and (\ref{sigmaQ})] as a function of $K$ at (a) $J_2=1.5$ and (b) $J_2=0.8$, using $L = 6,8$ cylinders.}
\label{sigma}
\end{figure}
%%%%%%%%%%%%%

%%%%%%%%%%%%%%%
\section{Nematicity}\label{sec:nematicity}

To characterize the nematicity in the different phases, we introduce two nematic order parameters $\sigma^S_{B1g}$ and $\sigma^Q_{B1g}$ defined as
%%%%%%%%%%%%%
\begin{eqnarray}
\sigma^S_{B1g} &=& \frac{1}{N_m}\sum_i[\langle {\bf S}_{i}\cdot {\bf S}_{i+\hat{x}}\rangle - \langle {\bf S}_{i}\cdot {\bf S}_{i+\hat{y}}\rangle] \label{sigma1}, \\
\sigma^Q_{B1g} &=& \frac{1}{N_m}\sum_i[\langle {\bf Q}_{i}\cdot {\bf Q}_{i+\hat{x}}\rangle - \langle {\bf Q}_{i}\cdot {\bf Q}_{i+\hat{y}}\rangle] \label{sigmaQ},
\end{eqnarray}
%%%%%%%%%%%%%
where $\hat{x}$ and $\hat{y}$ denote the unit length vectors along the $x$ and $y$ directions, respectively, and $N_m$ is the number of sites of the two columns in the middle of the cylinder. Analyses of these nematic order parameters have been shown efficient to determine the lattice rotational symmetry breaking in the DMRG calculations on the cylinder geometry~\cite{hu2020}. The absolute value of the nematic order parameters as a function of $K$ at $J_2 = 1.5$ and $0.8$ are, respectively, presented in Figs.~\ref{sigma}(a) and (b). Comparing $\sigma^S_{B1g}$ and $\sigma^Q_{B1g}$ clarifies whether the antiferromagnetic or antiquadrupolar fluctuations dominate the contributions to the nematic order. We find $\sigma^S_{B1g}$ dominating over $\sigma^Q_{B1g}$ inside the CAFM phase ($K \lesssim 0.65$ $(0.25)$ for $J_2 = 1.5$ $(0.8)$), and vice versa inside the $(\pi,0)$ AFQ phase  ($K \gtrsim  0.65$ for both values of $J_2$). We also notice that the crossings of the two nematic order parameters occurs exactly at the location where the quadrupolar order at ${\bf q}=(\pi,0)$ develops. This reflects the different types of fluctuations that are responsible for the nematic order on the two sides of the quantum phase transition. For $J_2 = 1.5$, the crossing is at the boundary between CAFM and $(\pi,0)$ AFQ, while for $J_2 = 0.8$ this crossing occurs at the boundary between CAFM and $(\pi/2,\pi)$ AFM$^*$.

%%%%%%%%%%%%%
\begin{figure}[t]
\begin{center}
\includegraphics[width=\columnwidth]{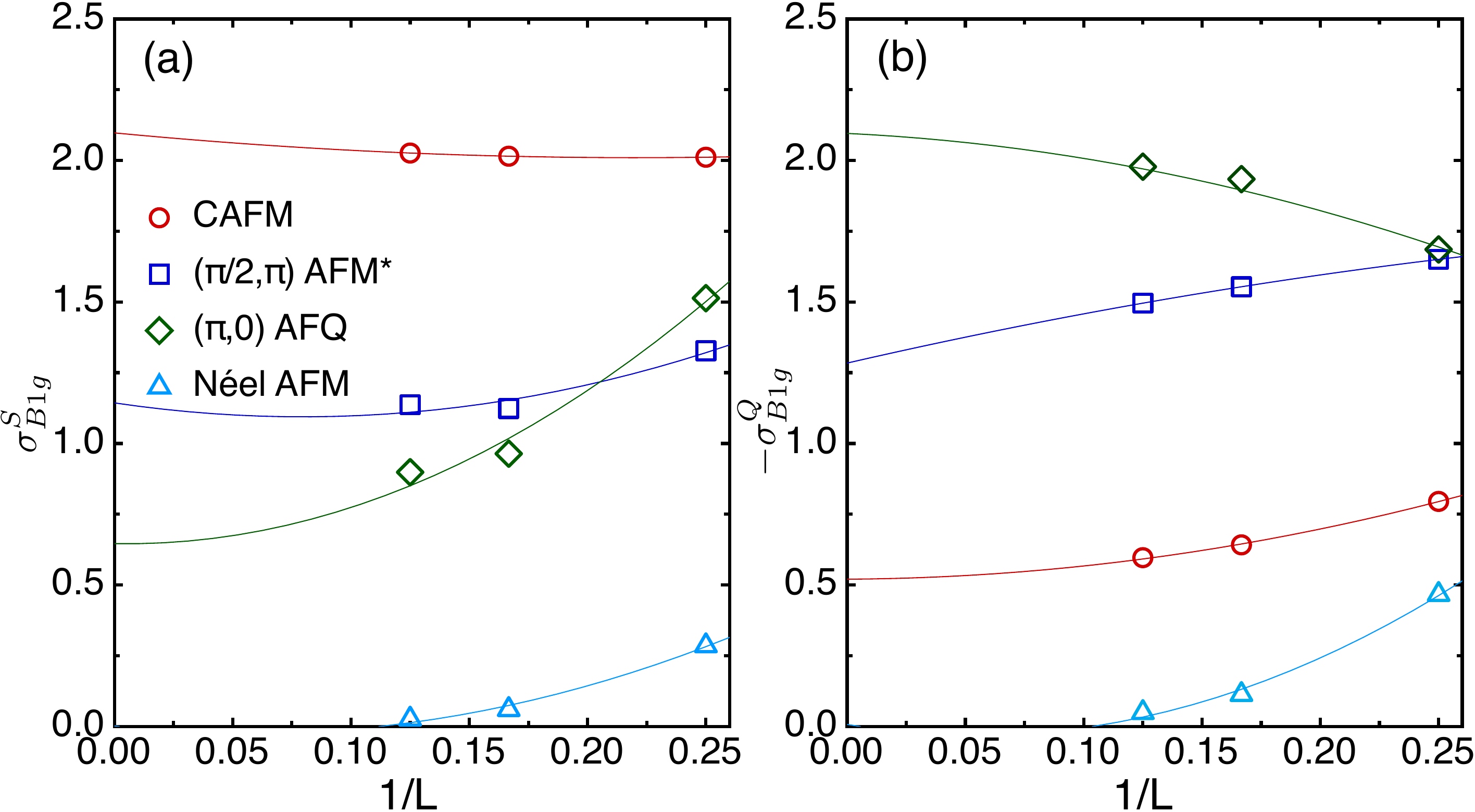}
\end{center}
\caption{The finite-size scaling of nematic order parameters $\sigma^S_{B1g}$ (a) and $\sigma^Q_{B1g}$ (b) for the N\'{e}el AFM state at $J_2=0.4$ and $K=0.6$, for the CAFM state at $J_2=1.5$ and  $K=0.5$, for the $(\pi,0)$ AFQ state at $J_2=1.5 $ and $K=0.7$, and for the ($\pi/2,\pi$) AFM$^*$ state at $J_2=0.8$ and $K=0.35$. The lines are guides to the eye.}
\label{nematicscaling}
\end{figure}
%%%%%%%%%%%%%

Finally, we show the finite-size scaling for the nematic order parameters $\sigma^S_{B1g}$ and $\sigma^Q_{B1g}$ of the four phases in Fig.~\ref{nematicscaling}. In the N\'{e}el AFM state, although there are small values of nematicity, due to the cylindric geometry used in the DMRG calculations, both $\sigma^S_{B1g}$ and $\sigma^Q_{B1g}$ decay fast and vanish with increasing size. For the other three phases, the finite-size scaling clearly indicates the presence of non-zero nematic orders in the thermodynamic limit.

%%%%%%%%%%%%%%%%%%%%%%%%%
\section{Discussions and Conclusions}\label{sec:dis}

\begin{figure}[t]
\begin{center}
\includegraphics[width=\columnwidth]{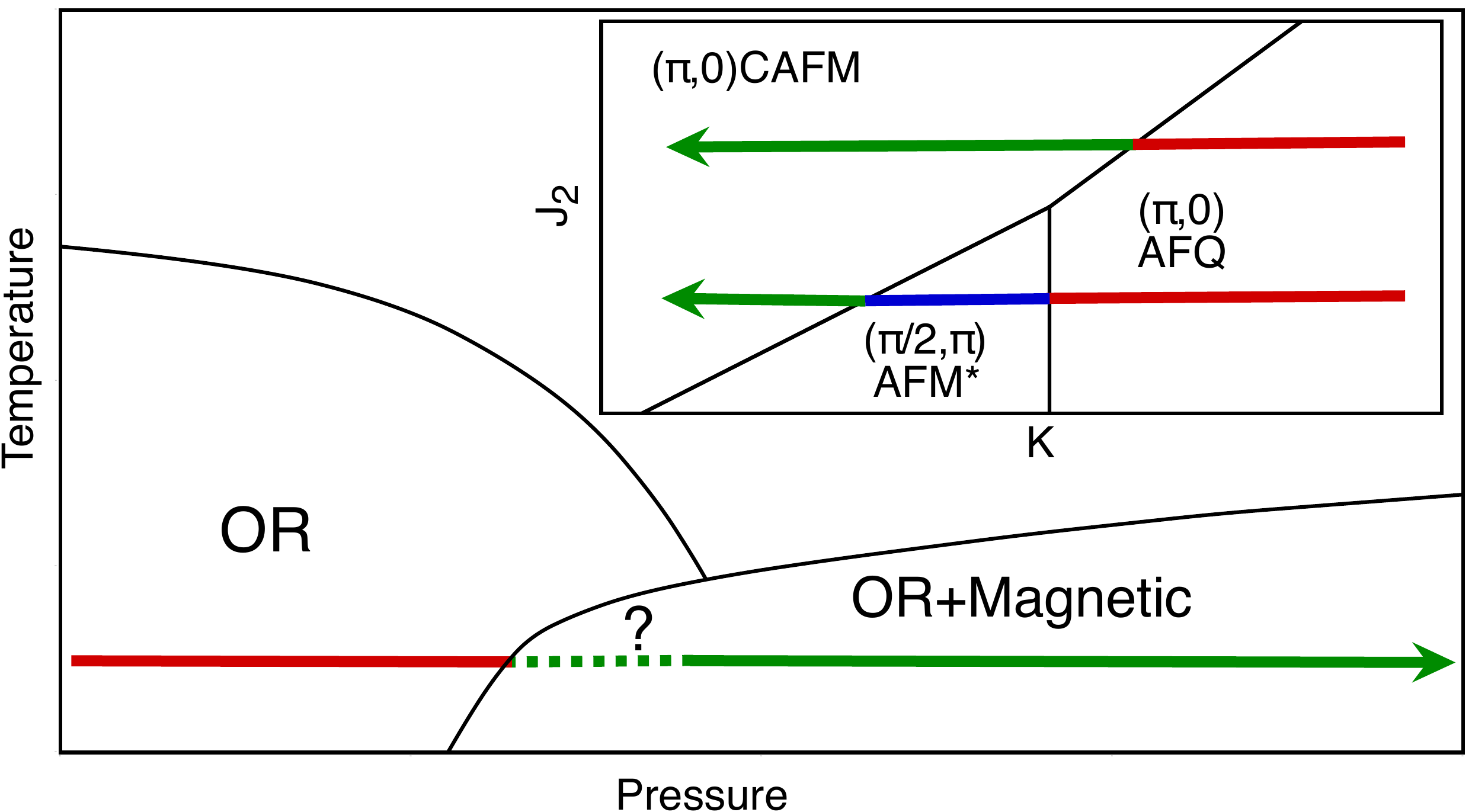}
\end{center}
\caption{The pressure-temperature phase diagram inferred from our theoretical phase diagram [illustrated in the inset, based on Fig.~\ref{mfpd}(a)]. There are two possible sequences of quantum phase transitions from the $(\pi,0)$ AFQ phase presumably stable at ambient pressure towards the high pressure CAFM $(\pi,0)$ phase, as illustrated by the arrows in both the main panel and the inset.}
\label{ptpd}
\end{figure}

We now discuss the implications of our results for the iron chalcogenides. Our work leads to a possible understanding of the properties of FeSe based on the presented phase diagram in Fig.~\ref{mfpd}(a). For clarity, we show a schematic phase diagram of the nematic phases in the inset of Fig.~\ref{ptpd}. Note that applying pressure increases the kinetic energy without affecting the local interactions as much and, thus, amounts to increasing $w$, the coherent electron spectral weight. Qualitatively, comparing the ambient-pressure FeSe with the pressurized FeSe is similar to comparing the ambient-pressure FeSe with the typical iron arsenides. Because the ambient-pressure FeSe is more strongly correlated than the latter, it is expected to be more frustrated and, correspondingly, having a larger $K/J$ ratio. In other words, under pressure, $K/J$ should decrease, which is the parameter trajectory we have proposed in Fig.~\ref{ptpd} for FeSe as a function of pressure. More microscopically, a non-perturbative procedure for calculating the effective exchange interactions in the bad metal regime has been developed using the slave-boson-type approach~\cite{ding2019}). Here, the bilinear spin-exchange interaction $J$ is given by a two-boson process and turns out to be $(1-w)$$J_c$, where $w$ denotes the percentage of the physical electron spectral weight that resides in the coherent part near the Fermi energy and $J_c$ is the exchange interaction at the delocalization-localization transition ({\it i.e.}, when $w\rightarrow0^+$). A similar procedure for the biquadratic interaction implies that it is given by a four-boson process, and will be on the order of $(1-w)^2$$K_c$. Thus, $K/J$ is expected to be proportional to $(1-w)$. Because applying pressure enhances the coherent electron spectral $w$, it is expected to lead to a decrease in $K/J$.

Thus, pressurizing FeSe may amount to taking a horizontal cut in this phase diagram: we propose two such cuts as candidates for the parameter tuning, which are also illustrated in the inset of Fig.~\ref{ptpd}. The resulting phase diagram is illustrated in the main panel of Fig.~\ref{ptpd}, with the system undergoing either a direct transition between the $(\pi,0)$ AFQ and CAFM states, or a transition between them through an intermediate $(\pi/2,\pi)$ AFM$^*$ regime with coexistence of magnetic and quadrupolar orders. This phase diagram is qualitatively similar to that inferred from recent experiments. While the presence of AFM order at pressures of the order of $2$~GPa had been indicated before,~\cite{Bendele2012} recent NMR measurements~\cite{Yuweiqiang2016} have provided strong evidence that the order achieved by increasing pressure breaks the $C_4$ symmetry and has a $(\pi,0)$ wave vector. The X-ray scattering experiments~\cite{Bohmer2016} have also provided evidence that a $C_4$ symmetry-breaking accompanies the magnetic ordering. There are indications in the existing experiments for two stages of phase transitions under pressure,~\cite{Miyoshi2014, Kaluarachchi2016,Bendele2012,Sun2015} with the onset of AFM order around $p_1 \approx 0.8$ GPa and a change of the magnetic structure around $p_2 \approx 1.2$ GPa.~\cite{Bendele2012} Additional NMR and neutron scattering measurements in the intermediate pressure range $0.8$ GPa $\lesssim P \lesssim $ $1.7$ GPa are especially needed to clarify this issue and ascertain which of the two proposed sequences applies. We reiterate that the set of model parameters we choose is for the purpose of illustrating the phase transitions in Fig.~\ref{ptpd}. The nematic phases and the transitions actually span over a large parameter regime in the overall phase diagram, which underscores the fact that the proposed physical picture is robust instead of fine-tuned.

%%%%%%%%%%%%%
%%%%%%%%%%%%%

Regardless of which of the two phase transition sequences is realized, our results have important implications for the single-electron excitations. The $(\pi,0)$ AFQ state contains two order parameters. The rank-$2$ AFQ order parameter does not efficiently couple with the (coherent) conduction electrons near the Fermi surface and, therefore, will not cause a reconstruction of the Fermi surface. However, the nematic order parameter, $\sigma^S_{B1g}$ and $\sigma^Q_{B1g}$ defined in Eqs.~(\ref{sigma1}) and (\ref{sigmaQ}), will linearly mix with the occupancy difference in the $3d_{xz}$ and $3d_{yz}$ orbitals, thereby generating a splitting of the electronic bands and a distortion of the Fermi surface. All these features are consistent with the observations by the ARPES experiments.~\cite{Nakayama2014,Shimojima2014,Watson2015a,Yi2019} Likewise, the $(0,\pi)$ CAFM state contains two order parameters. While the nematic order parameter acts similarly as in the AFQ case, distorting the Fermi surface, the AFM order is very different from the AFQ order: it provides a spatially modulated and spin-dependent potential for the conduction electrons, thereby reconstructing the Fermi surface. Thus, our proposed quantum phase transitions will be accompanied by drastic changes in the geometry of the Fermi surfaces. This is consistent with the dramatic evolution of the Fermi surface recently reported in the Shubnikov-de Haas (SdH) oscillation measurements on FeSe.~\cite{Terashima2016}

To summarize, the present work has advanced two key results. First, by considering the interplay between the frustrated magnetic interactions, we establish a quantum phase diagram in which the (non-magnetic) ($\pi$,0) antiferroquadrupolar order is {\it robustly} located near the $(\pi,0)$ collinear antiferromagnetic order and an intervening $(\pi/2,\pi)$ antiferromagnetic phase, all of which break the C$_4$ symmetry and thus promote a nematic order. Second, this theoretical result provides the basis to understand qualitatively the quantum phase transition of FeSe under pressure. The similarity of the quantum phase transitions we have identified in the frustrated bilinear-biquadratic model with the experimental observations provides evidence that a similar type of spin physics is important for the emergence of superconductivity in both iron chalcogenides and iron pnictides. This is not to say that the orbital degrees of freedom are decoupled. As discussed above, the nematic order of the spin quadrupolar or dipolar orders will be coupled to the orbital order. Nonetheless, the interactions among the spin degrees of freedom, as described in Eq.~(\ref{ham}) %Eqs.~(\ref{sigma1}) and (\ref{sigmaQ}), 
will give rise to superconducting pairing in FeSe -- and by extension in other iron chalcogenides -- in a similar way as they do in the iron pnictides. Thus, our results not only contribute to the understanding of recent experiments in FeSe, but also provide evidence for a common origin of superconductivity across the extensive material classes of iron-based superconductors. More generally, our findings suggest the importance of correlation-induced short-range spin exchange interactions for both the normal state and superconductivity in the iron chalcogenides. This provides a new linkage between the superconductivity of the highest $T_c$ iron-based families with that arising in a broad array of strongly correlated electron systems, including the cuprates and heavy fermion metals.
\\

%%%%%%%%%%%%%
\begin{center}
{\bf ACKNOWLEDGMENTS}
\end{center}

We thank W. Yu, A. Goldman, E. Abrahams, A. B\"ohmer, A. Boothroyd, A. Coldea, A. H. Nevidomskyy, and E. M. Nica for useful discussions. E.D. and W.-J.H. were supported by the U.S. Department of Energy (DOE), Office of Science, Basic Energy Sciences (BES),  Materials  Science  and  Engineering  Division. This work was supported in part by the U.S. Department of Energy, Office of Science, Basic Energy Sciences, under Award No. DE-SC0018197 and the Robert A.\ Welch Foundation Grant No.\ C-1411 (W.-J.H., H.-H.L. and Q.S.), a Smalley Postdoctoral Fellowship of the Rice Center for Quantum Materials (H-H. L.).
R. Y. was supported by the National Science Foundation of China Grant number 11674392, the Fundamental Research Funds for the Central Universities and the Research Funds of Remnin University of China Grant number 18XNLG24, and the Ministry of Science and Technology of China, National Program on Key Research Project Grant number 2016YFA0300504.
S.S.G. was supported by NSFC grants No. 11874078, 11834014, and the Fundamental Research Funds for the Central Universities.
The majority of the computational calculations have been performed on the Extreme Science and Engineering Discovery Environment (XSEDE) supported by NSF under Grant No.\ DMR160057. Most of the numerical calculations have been done by W.-J.H. and H.-H.L. while at Rice University.

%%%%%%
\appendix

\section{Additional DMRG Phase Diagrams}\label{app1}

\begin{figure*}[tbp]
\begin{center}
\includegraphics[width=1.6\columnwidth]{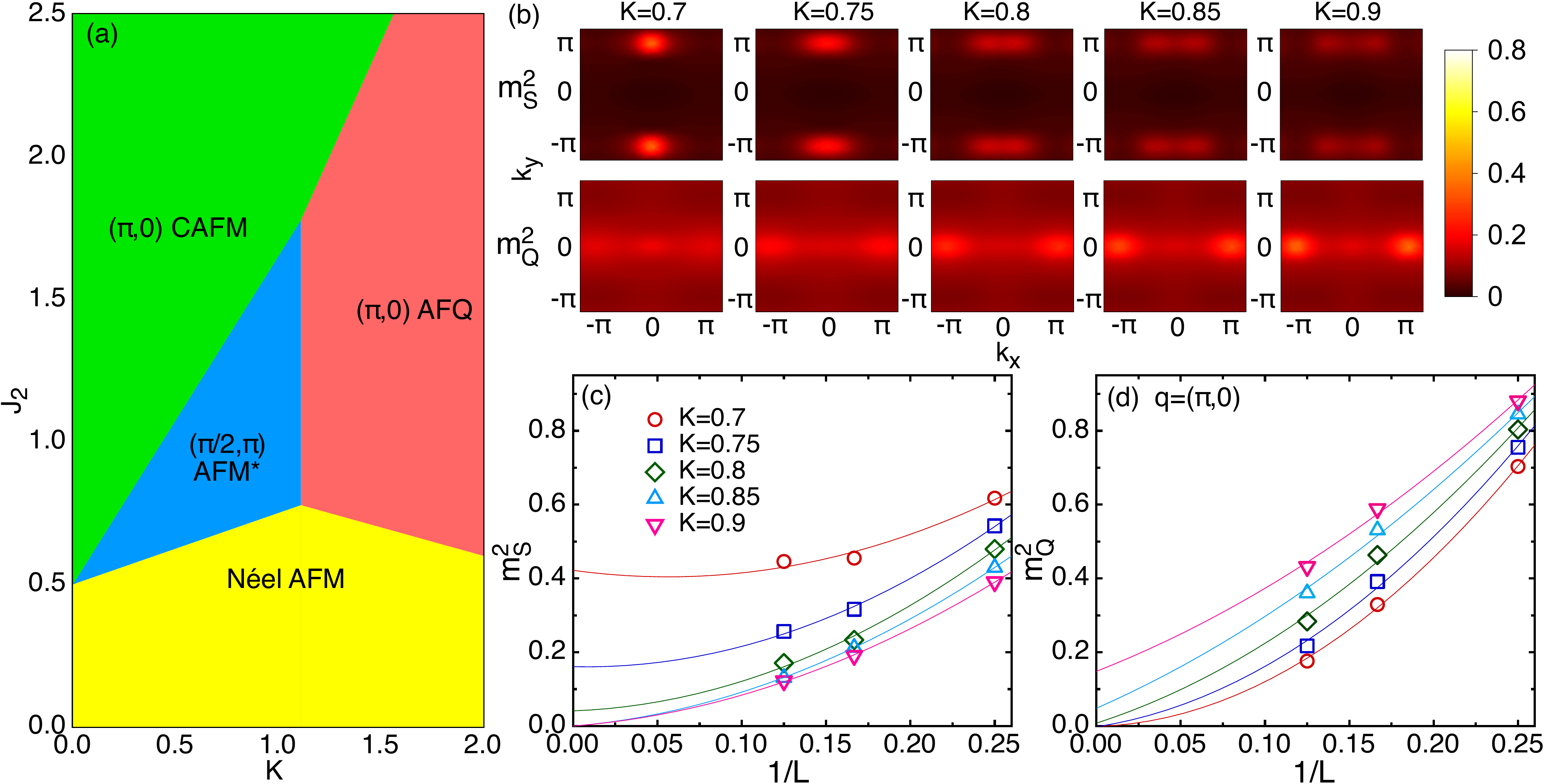}
\end{center}
\caption{(a) Phase diagram of the $J_1$-$J_2$-$K_1$-$K_2$-$K_3$ model with $K_1=-K, K_2=0.9K, K_3=-0.7K$ ($K>0$) on the $J_2$-$K$ plane ($J_1$ is set to $1.0$). The phase boundaries are determined from site-factorized wave-function calculations. (b) Spin ($m^2_{S}$) and quadrupolar ($m^2_{Q}$) structure factors obtained from DMRG calculations using $8\times 16$ cylinders for $J_2=1.5$ with $K_1=-K, K_2=0.9K, K_3=-0.7K$. (c) Finite-size scaling for the spin structure factor, at the highest peak of $m^{2}_S$ in its momentum distribution, is shown. (d) Finite-size scaling for the quadrupolar structure factors with the intensity at ${\bf q}=(\pi,0)$ being plotted. According to the scaling, $K=0.8$ is close to the phase boundary. The lines are guides to the eye.}
\label{pd97}
\end{figure*}

\begin{figure*}[tbp]
\begin{center}
\includegraphics[width=1.6\columnwidth]{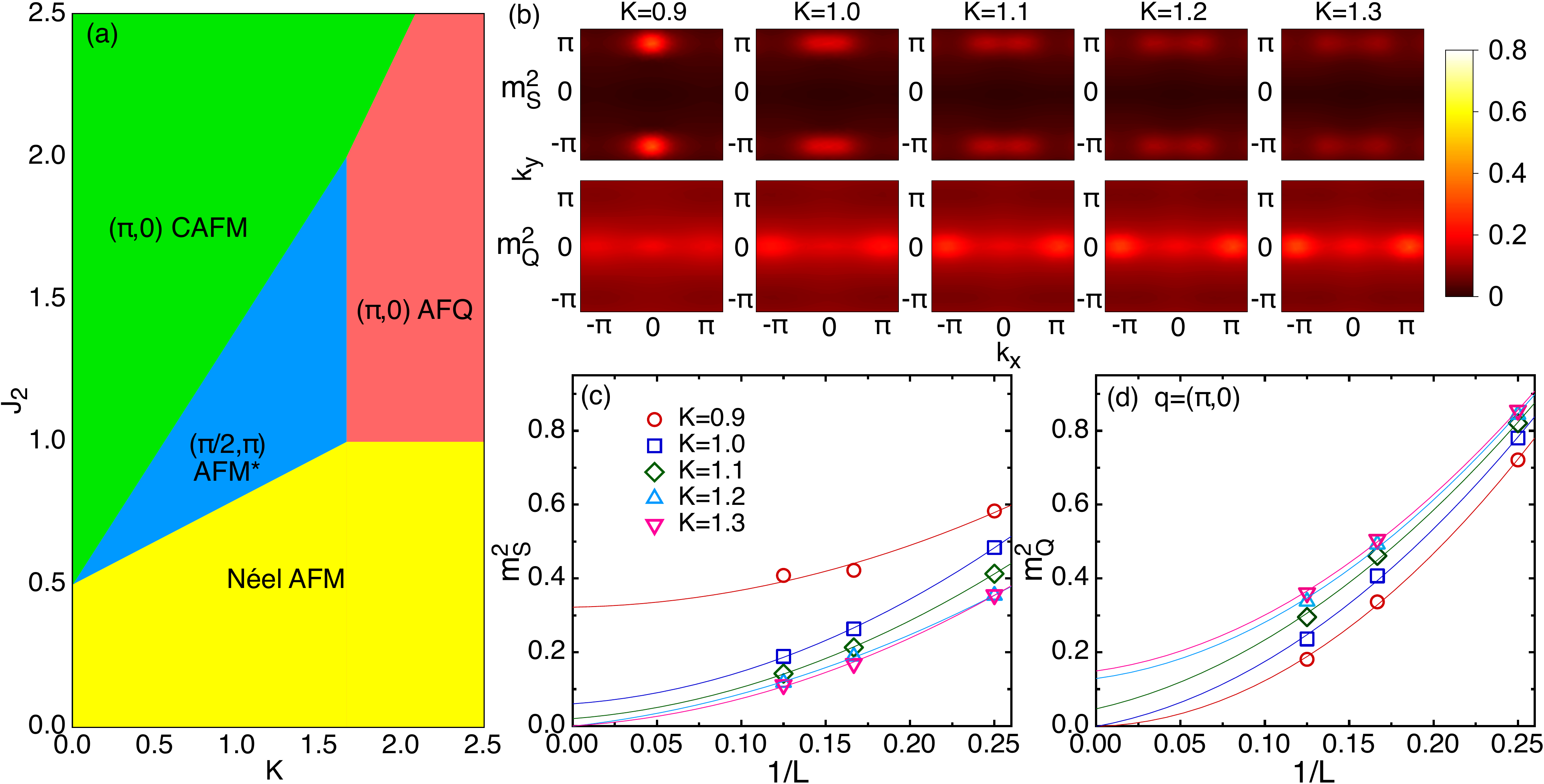}
\end{center}
\caption{(a) Phase diagram of the $J_1$-$J_2$-$K_1$-$K_2$-$K_3$ model with $K_1=-K, K_2=0.7K, K_3=-0.5K$ ($K>0$) on the $J_2$-$K$ plane ($J_1$ is set to $1.0$). The phase boundaries are determined from site-factorized wave-function calculations. (b) Spin ($m^2_{S}$) and quadrupolar ($m^2_{Q}$) structure factors obtained from DMRG calculations using $8\times 16$ cylinders for $J_2=1.5$ with $K_1=-K, K_2=0.7K, K_3=-0.5K$. (c) Finite-size scaling for the spin structure factor, at the highest peak of $m^{2}_S$ in its momentum distribution, is shown. (d) Finite-size scaling for the quadrupolar structure factors with the intensity at ${\bf q}=(\pi,0)$ being plotted. According to the scaling, $K=1.1$ is close to the phase boundary. The lines are guides to the eye.}
\label{pd75}
\end{figure*}

We have performed the site-factorized wave-function analysis and DMRG simulations for additions parameter sets, with results for $(K_1=-K, K_2=0.9K, K_3=-0.7K)$ shown in Fig.~\ref{pd97} and for $(K_1=-K, K_2=0.7K, K_3=-0.5K)$ shown in Fig.~\ref{pd75}. In both cases, we have chosen $J_1=1$ and $J_2=1.5$ for the DMRG calculations. From the finite-size scalings of the spin ($m^2_S$) and quadrupolar ($m^2_Q$) order parameters, we find direct phase transitions between the CAFM and $(\pi,0)$ AFQ phases for both cases, and the transition points are around $K=0.8$ for the parameter set $(K_1=-K, K_2=0.9K, K_3=-0.7K)$ and $K=1.1$ for $(K_1=-K, K_2=0.7K, K_3=-0.5K)$. These results suggest that the nematic phases actually span over large parameter regions in the phase diagram, and the general features of the nematic phase diagram and relation to experiments remain qualitatively the same as described in our conclusions of the main text.

%%%%%%%%%%%%%
%%%%%%%%%%%%%

%%%%%%%%%%%%%
%%%%%%%%%%%%%

\section{Ground-State Energy in the Site-Factorized Wave-Function Approximation}\label{app2}

Within the site-factorized wave-function approximation of the $SU(3)$ representation, the ground-state energy per site of a certain ordered phase in the $S=1$ model can be readily determined. Here we show how this works for the $(\pi,0)$ CAFM state, with particular emphasis on the quantum contributions. The generalization to other states is straightforward.

Denote the two sublattices of the CAFM state to be $A$ and $B$, respectively. Without losing generality, we assume the local wave functions on these two sublattices to be
\begin{eqnarray}
 |\Psi_A \rangle = |1\rangle = \frac{1}{\sqrt{2}} \left( -i|x\rangle +|y\rangle \right), \nonumber \\
 |\Psi_B \rangle = |-1\rangle = \frac{1}{\sqrt{2}} \left( i|x\rangle +|y\rangle \right).
\end{eqnarray} 
The corresponding directors are, respectively, 
\begin{eqnarray}
 {\bf d}_A = (-\frac{i}{\sqrt{2}}, \frac{1}{\sqrt{2}}, 0), \nonumber\\
{\bf d}_B = (\frac{i}{\sqrt{2}}, \frac{1}{\sqrt{2}}, 0).
\end{eqnarray}
This gives 
\begin{eqnarray}
 {\bf d}_A \cdot {\bf d}_B =1, \quad {\bf d}_A \cdot \bar{{\bf d}}_B =0
\end{eqnarray}
for the antiferromagnetically coupled bond, and 
\begin{eqnarray}
 {\bf d}_A \cdot {\bf d}_A =0, \quad {\bf d}_A \cdot \bar{{\bf d}}_A =1
\end{eqnarray}
for the ferromagnetic coupled bond. For the CAFM state, a site connects to one AFM and one FM nearest-neighbor bonds, two AFM next-nearest-neighbor bonds, and two FM 3rd-nearest-neighbor bonds on average. Following Eq.~\eqref{Eq:H_d}, one gets the energy per site for the CAFM phase to be
\begin{eqnarray}
 \mathcal{E}_{CAFM} = -2J_2 + 2J_3 + K_1 + 2K_2  = -2J_2+K,
\end{eqnarray}
when neglecting the constant term $K_{ij}$ in Eq.~\eqref{Eq:H_d}. The contribution from the constant term is $2(K_1+K_2+K_3)=-2K$, and this shifts the energy to $\mathcal{E}^\prime_{CAFM} = -2J_2 + 2J_3 + 3K_1 + 4K_2 + 2K_3 = -2J_2-K$.

Note that this energy is higher than the energy per site of the CAFM state in the classical limit, $\mathcal{E}^{c}_{CAFM} = -2J_2 + 2J_3 + 2K_1 + 2K_2 + 2K_3 = -2J_2-2K$. The reason is as follows. In the classical limit, $\langle ({\bf S}_A\cdot {\bf S}_B)^2\rangle = \langle (S^z_A S^z_B)^2 \rangle = 1$ for an AFM bond. But in the $SU(3)$ representation of the $S=1$ model, one can show that for the AFM bond $\langle ({\bf S}_A\cdot {\bf S}_B)^2\rangle = \langle (S^z_A S^z_B)^2 + S^+_A S^-_B S^-_A S^+_B/4\rangle = 2$. The larger value comes from the transverse correlation $S^+_A S^-_B S^-_A S^+_B/4$ and reflects the inherent quantum mechanical nature of the AFM state. Note that there are one nearest neighbor and two next-nearest-neighbor AFM bonds in the CAFM state, therefore the energy difference between the $S=1$ case and the classical limit is $\mathcal{E}^\prime_{CAFM} - \mathcal{E}^{c}_{CAFM} = K_1 + 2K_2 = K$. By contrast, for any FM bond, the transverse corrections vanish.

\clearpage
%%%%%%
%\bibliographystyle{apsrev}
\bibliography{qpt}

\begin{thebibliography}{50}%
\makeatletter
\providecommand \@ifxundefined [1]{%
 \@ifx{#1\undefined}
}%
\providecommand \@ifnum [1]{%
 \ifnum #1\expandafter \@firstoftwo
 \else \expandafter \@secondoftwo
 \fi
}%
\providecommand \@ifx [1]{%
 \ifx #1\expandafter \@firstoftwo
 \else \expandafter \@secondoftwo
 \fi
}%
\providecommand \natexlab [1]{#1}%
\providecommand \enquote  [1]{``#1''}%
\providecommand \bibnamefont  [1]{#1}%
\providecommand \bibfnamefont [1]{#1}%
\providecommand \citenamefont [1]{#1}%
\providecommand \href@noop [0]{\@secondoftwo}%
\providecommand \href [0]{\begingroup \@sanitize@url \@href}%
\providecommand \@href[1]{\@@startlink{#1}\@@href}%
\providecommand \@@href[1]{\endgroup#1\@@endlink}%
\providecommand \@sanitize@url [0]{\catcode `\\12\catcode `\$12\catcode
  `\&12\catcode `\#12\catcode `\^12\catcode `\_12\catcode `\%12\relax}%
\providecommand \@@startlink[1]{}%
\providecommand \@@endlink[0]{}%
\providecommand \url  [0]{\begingroup\@sanitize@url \@url }%
\providecommand \@url [1]{\endgroup\@href {#1}{\urlprefix }}%
\providecommand \urlprefix  [0]{URL }%
\providecommand \Eprint [0]{\href }%
\providecommand \doibase [0]{http://dx.doi.org/}%
\providecommand \selectlanguage [0]{\@gobble}%
\providecommand \bibinfo  [0]{\@secondoftwo}%
\providecommand \bibfield  [0]{\@secondoftwo}%
\providecommand \translation [1]{[#1]}%
\providecommand \BibitemOpen [0]{}%
\providecommand \bibitemStop [0]{}%
\providecommand \bibitemNoStop [0]{.\EOS\space}%
\providecommand \EOS [0]{\spacefactor3000\relax}%
\providecommand \BibitemShut  [1]{\csname bibitem#1\endcsname}%
\let\auto@bib@innerbib\@empty
%</preamble>
\bibitem [{\citenamefont {Kamihara}\ \emph {et~al.}(2008)\citenamefont
  {Kamihara}, \citenamefont {Watanabe}, \citenamefont {Hirano},\ and\
  \citenamefont {Hosono}}]{Kamihara2008}%
  \BibitemOpen
  \bibfield  {author} {\bibinfo {author} {\bibfnamefont {Yoichi}\ \bibnamefont
  {Kamihara}}, \bibinfo {author} {\bibfnamefont {Takumi}\ \bibnamefont
  {Watanabe}}, \bibinfo {author} {\bibfnamefont {Masahiro}\ \bibnamefont
  {Hirano}}, \ and\ \bibinfo {author} {\bibfnamefont {Hideo}\ \bibnamefont
  {Hosono}},\ }\bibfield  {title} {\enquote {\bibinfo {title} {Iron-based
  layered superconductor lala[o$_{1-x}$f$_x$]feas ($x$ = $0.05$-$0.12$) with
  $t_c$ = $26$ k},}\ }\href {https://doi.org/10.1021/ja800073m} {\bibfield
  {journal} {\bibinfo  {journal} {Journal of the American Chemical Society}\
  }\textbf {\bibinfo {volume} {130}},\ \bibinfo {pages} {3296--3297} (\bibinfo
  {year} {2008})}\BibitemShut {NoStop}%
\bibitem [{\citenamefont {Stewart}(2011)}]{Stewart2011}%
  \BibitemOpen
  \bibfield  {author} {\bibinfo {author} {\bibfnamefont {G.~R.}\ \bibnamefont
  {Stewart}},\ }\bibfield  {title} {\enquote {\bibinfo {title}
  {Superconductivity in iron compounds},}\ }\href {\doibase
  10.1103/RevModPhys.83.1589} {\bibfield  {journal} {\bibinfo  {journal} {Rev.
  Mod. Phys.}\ }\textbf {\bibinfo {volume} {83}},\ \bibinfo {pages}
  {1589--1652} (\bibinfo {year} {2011})}\BibitemShut {NoStop}%
\bibitem [{\citenamefont {Si}\ \emph {et~al.}(2016)\citenamefont {Si},
  \citenamefont {Yu},\ and\ \citenamefont {Abrahams}}]{Si2016}%
  \BibitemOpen
  \bibfield  {author} {\bibinfo {author} {\bibfnamefont {Qimiao}\ \bibnamefont
  {Si}}, \bibinfo {author} {\bibfnamefont {Rong}\ \bibnamefont {Yu}}, \ and\
  \bibinfo {author} {\bibfnamefont {Elihu}\ \bibnamefont {Abrahams}},\
  }\bibfield  {title} {\enquote {\bibinfo {title} {High-temperature
  superconductivity in iron pnictides and chalcogenides},}\ }\href
  {https://doi.org/10.1038/natrevmats.2016.17} {\bibfield  {journal} {\bibinfo
  {journal} {Nature Reviews Materials}\ }\textbf {\bibinfo {volume} {1}},\
  \bibinfo {pages} {16017} (\bibinfo {year} {2016})}\BibitemShut {NoStop}%
\bibitem [{\citenamefont {Dai}\ \emph {et~al.}(2012)\citenamefont {Dai},
  \citenamefont {Hu},\ and\ \citenamefont {Dagotto}}]{PCDai_review12}%
  \BibitemOpen
  \bibfield  {author} {\bibinfo {author} {\bibfnamefont {Pengcheng}\
  \bibnamefont {Dai}}, \bibinfo {author} {\bibfnamefont {Jiangping}\
  \bibnamefont {Hu}}, \ and\ \bibinfo {author} {\bibfnamefont {Elbio}\
  \bibnamefont {Dagotto}},\ }\bibfield  {title} {\enquote {\bibinfo {title}
  {Magnetism and its microscopic origin in iron-based high-temperature
  superconductors},}\ }\href {https://doi.org/10.1038/nphys2438} {\bibfield
  {journal} {\bibinfo  {journal} {Nature Physics}\ }\textbf {\bibinfo {volume}
  {8}},\ \bibinfo {pages} {709} (\bibinfo {year} {2012})}\BibitemShut {NoStop}%
\bibitem [{\citenamefont {Dagotto}(2013)}]{Elbio_rmp}%
  \BibitemOpen
  \bibfield  {author} {\bibinfo {author} {\bibfnamefont {Elbio}\ \bibnamefont
  {Dagotto}},\ }\bibfield  {title} {\enquote {\bibinfo {title} {Colloquium: The
  unexpected properties of alkali metal iron selenide superconductors},}\
  }\href {\doibase 10.1103/RevModPhys.85.849} {\bibfield  {journal} {\bibinfo
  {journal} {Rev. Mod. Phys.}\ }\textbf {\bibinfo {volume} {85}},\ \bibinfo
  {pages} {849--867} (\bibinfo {year} {2013})}\BibitemShut {NoStop}%
\bibitem [{\citenamefont {Guo}\ \emph {et~al.}(2010)\citenamefont {Guo},
  \citenamefont {Jin}, \citenamefont {Wang}, \citenamefont {Wang},
  \citenamefont {Zhu}, \citenamefont {Zhou}, \citenamefont {He},\ and\
  \citenamefont {Chen}}]{Guo2010}%
  \BibitemOpen
  \bibfield  {author} {\bibinfo {author} {\bibfnamefont {Jiangang}\
  \bibnamefont {Guo}}, \bibinfo {author} {\bibfnamefont {Shifeng}\ \bibnamefont
  {Jin}}, \bibinfo {author} {\bibfnamefont {Gang}\ \bibnamefont {Wang}},
  \bibinfo {author} {\bibfnamefont {Shunchong}\ \bibnamefont {Wang}}, \bibinfo
  {author} {\bibfnamefont {Kaixing}\ \bibnamefont {Zhu}}, \bibinfo {author}
  {\bibfnamefont {Tingting}\ \bibnamefont {Zhou}}, \bibinfo {author}
  {\bibfnamefont {Meng}\ \bibnamefont {He}}, \ and\ \bibinfo {author}
  {\bibfnamefont {Xiaolong}\ \bibnamefont {Chen}},\ }\bibfield  {title}
  {\enquote {\bibinfo {title} {Superconductivity in the iron selenide
  ${\text{k}}_{x}{\text{fe}}_{2}{\text{se}}_{2}$
  $(0\ensuremath{\le}x\ensuremath{\le}1.0)$},}\ }\href {\doibase
  10.1103/PhysRevB.82.180520} {\bibfield  {journal} {\bibinfo  {journal} {Phys.
  Rev. B}\ }\textbf {\bibinfo {volume} {82}},\ \bibinfo {pages} {180520}
  (\bibinfo {year} {2010})}\BibitemShut {NoStop}%
\bibitem [{\citenamefont {Wang}\ \emph {et~al.}(2012)\citenamefont {Wang},
  \citenamefont {Li}, \citenamefont {Zhang}, \citenamefont {Zhang},
  \citenamefont {Zhang}, \citenamefont {Li}, \citenamefont {Ding},
  \citenamefont {Ou}, \citenamefont {Deng}, \citenamefont {Chang},
  \citenamefont {Wen}, \citenamefont {Song}, \citenamefont {He}, \citenamefont
  {Jia}, \citenamefont {Ji}, \citenamefont {Wang}, \citenamefont {Wang},
  \citenamefont {Chen}, \citenamefont {Ma},\ and\ \citenamefont
  {Xue}}]{QYWang2012}%
  \BibitemOpen
  \bibfield  {author} {\bibinfo {author} {\bibfnamefont {Qing-Yan}\
  \bibnamefont {Wang}}, \bibinfo {author} {\bibfnamefont {Zhi}\ \bibnamefont
  {Li}}, \bibinfo {author} {\bibfnamefont {Wen-Hao}\ \bibnamefont {Zhang}},
  \bibinfo {author} {\bibfnamefont {Zuo-Cheng}\ \bibnamefont {Zhang}}, \bibinfo
  {author} {\bibfnamefont {Jin-Song}\ \bibnamefont {Zhang}}, \bibinfo {author}
  {\bibfnamefont {Wei}\ \bibnamefont {Li}}, \bibinfo {author} {\bibfnamefont
  {Hao}\ \bibnamefont {Ding}}, \bibinfo {author} {\bibfnamefont {Yun-Bo}\
  \bibnamefont {Ou}}, \bibinfo {author} {\bibfnamefont {Peng}\ \bibnamefont
  {Deng}}, \bibinfo {author} {\bibfnamefont {Kai}\ \bibnamefont {Chang}},
  \bibinfo {author} {\bibfnamefont {Jing}\ \bibnamefont {Wen}}, \bibinfo
  {author} {\bibfnamefont {Can-Li}\ \bibnamefont {Song}}, \bibinfo {author}
  {\bibfnamefont {Ke}~\bibnamefont {He}}, \bibinfo {author} {\bibfnamefont
  {Jin-Feng}\ \bibnamefont {Jia}}, \bibinfo {author} {\bibfnamefont
  {Shuai-Hua}\ \bibnamefont {Ji}}, \bibinfo {author} {\bibfnamefont {Ya-Yu}\
  \bibnamefont {Wang}}, \bibinfo {author} {\bibfnamefont {Li-Li}\ \bibnamefont
  {Wang}}, \bibinfo {author} {\bibfnamefont {Xi}~\bibnamefont {Chen}}, \bibinfo
  {author} {\bibfnamefont {Xu-Cun}\ \bibnamefont {Ma}}, \ and\ \bibinfo
  {author} {\bibfnamefont {Qi-Kun}\ \bibnamefont {Xue}},\ }\bibfield  {title}
  {\enquote {\bibinfo {title} {Interface-induced high-temperature
  superconductivity in single unit-cell {FeSe} films on {SrTiO}3},}\ }\href
  {\doibase 10.1088/0256-307x/29/3/037402} {\bibfield  {journal} {\bibinfo
  {journal} {Chinese Physics Letters}\ }\textbf {\bibinfo {volume} {29}},\
  \bibinfo {pages} {037402} (\bibinfo {year} {2012})}\BibitemShut {NoStop}%
\bibitem [{\citenamefont {Lee}\ \emph {et~al.}(2014)\citenamefont {Lee},
  \citenamefont {Schmitt}, \citenamefont {Moore}, \citenamefont {Johnston},
  \citenamefont {Cui}, \citenamefont {Li}, \citenamefont {Yi}, \citenamefont
  {Liu}, \citenamefont {Hashimoto}, \citenamefont {Zhang}, \citenamefont {Lu},
  \citenamefont {Devereaux}, \citenamefont {Lee},\ and\ \citenamefont
  {Shen}}]{JJLee_Nature}%
  \BibitemOpen
  \bibfield  {author} {\bibinfo {author} {\bibfnamefont {J.~J.}\ \bibnamefont
  {Lee}}, \bibinfo {author} {\bibfnamefont {F.~T.}\ \bibnamefont {Schmitt}},
  \bibinfo {author} {\bibfnamefont {R.~G.}\ \bibnamefont {Moore}}, \bibinfo
  {author} {\bibfnamefont {S.}~\bibnamefont {Johnston}}, \bibinfo {author}
  {\bibfnamefont {Y.~T.}\ \bibnamefont {Cui}}, \bibinfo {author} {\bibfnamefont
  {W.}~\bibnamefont {Li}}, \bibinfo {author} {\bibfnamefont {M.}~\bibnamefont
  {Yi}}, \bibinfo {author} {\bibfnamefont {Z.~K.}\ \bibnamefont {Liu}},
  \bibinfo {author} {\bibfnamefont {M.}~\bibnamefont {Hashimoto}}, \bibinfo
  {author} {\bibfnamefont {Y.}~\bibnamefont {Zhang}}, \bibinfo {author}
  {\bibfnamefont {D.~H.}\ \bibnamefont {Lu}}, \bibinfo {author} {\bibfnamefont
  {T.~P.}\ \bibnamefont {Devereaux}}, \bibinfo {author} {\bibfnamefont {D.~H.}\
  \bibnamefont {Lee}}, \ and\ \bibinfo {author} {\bibfnamefont {Z.~X.}\
  \bibnamefont {Shen}},\ }\bibfield  {title} {\enquote {\bibinfo {title}
  {Interfacial mode coupling as the origin of the enhancement of tc in fese
  films on srtio3},}\ }\href {\doibase 10.1038/nature13894} {\bibfield
  {journal} {\bibinfo  {journal} {Nature}\ }\textbf {\bibinfo {volume} {515}},\
  \bibinfo {pages} {245--248} (\bibinfo {year} {2014})}\BibitemShut {NoStop}%
\bibitem [{\citenamefont {Hsu}\ \emph {et~al.}(2008)\citenamefont {Hsu},
  \citenamefont {Luo}, \citenamefont {Yeh}, \citenamefont {Chen}, \citenamefont
  {Huang}, \citenamefont {Wu}, \citenamefont {Lee}, \citenamefont {Huang},
  \citenamefont {Chu}, \citenamefont {Yan},\ and\ \citenamefont {Wu}}]{Wu08}%
  \BibitemOpen
  \bibfield  {author} {\bibinfo {author} {\bibfnamefont {Fong-Chi}\
  \bibnamefont {Hsu}}, \bibinfo {author} {\bibfnamefont {Jiu-Yong}\
  \bibnamefont {Luo}}, \bibinfo {author} {\bibfnamefont {Kuo-Wei}\ \bibnamefont
  {Yeh}}, \bibinfo {author} {\bibfnamefont {Ta-Kun}\ \bibnamefont {Chen}},
  \bibinfo {author} {\bibfnamefont {Tzu-Wen}\ \bibnamefont {Huang}}, \bibinfo
  {author} {\bibfnamefont {Phillip~M.}\ \bibnamefont {Wu}}, \bibinfo {author}
  {\bibfnamefont {Yong-Chi}\ \bibnamefont {Lee}}, \bibinfo {author}
  {\bibfnamefont {Yi-Lin}\ \bibnamefont {Huang}}, \bibinfo {author}
  {\bibfnamefont {Yan-Yi}\ \bibnamefont {Chu}}, \bibinfo {author}
  {\bibfnamefont {Der-Chung}\ \bibnamefont {Yan}}, \ and\ \bibinfo {author}
  {\bibfnamefont {Maw-Kuen}\ \bibnamefont {Wu}},\ }\bibfield  {title} {\enquote
  {\bibinfo {title} {Superconductivity in the pbo-type structure
  $\alpha$-fese},}\ }\href {\doibase 10.1073/pnas.0807325105} {\bibfield
  {journal} {\bibinfo  {journal} {Proceedings of the National Academy of
  Sciences}\ }\textbf {\bibinfo {volume} {105}},\ \bibinfo {pages}
  {14262--14264} (\bibinfo {year} {2008})}\BibitemShut {NoStop}%
\bibitem [{\citenamefont {Fang}\ \emph {et~al.}(2008)\citenamefont {Fang},
  \citenamefont {Pham}, \citenamefont {Qian}, \citenamefont {Liu},
  \citenamefont {Vehstedt}, \citenamefont {Liu}, \citenamefont {Spinu},\ and\
  \citenamefont {Mao}}]{Mao08}%
  \BibitemOpen
  \bibfield  {author} {\bibinfo {author} {\bibfnamefont {M.~H.}\ \bibnamefont
  {Fang}}, \bibinfo {author} {\bibfnamefont {H.~M.}\ \bibnamefont {Pham}},
  \bibinfo {author} {\bibfnamefont {B.}~\bibnamefont {Qian}}, \bibinfo {author}
  {\bibfnamefont {T.~J.}\ \bibnamefont {Liu}}, \bibinfo {author} {\bibfnamefont
  {E.~K.}\ \bibnamefont {Vehstedt}}, \bibinfo {author} {\bibfnamefont
  {Y.}~\bibnamefont {Liu}}, \bibinfo {author} {\bibfnamefont {L.}~\bibnamefont
  {Spinu}}, \ and\ \bibinfo {author} {\bibfnamefont {Z.~Q.}\ \bibnamefont
  {Mao}},\ }\bibfield  {title} {\enquote {\bibinfo {title} {Superconductivity
  close to magnetic instability in
  $\text{Fe}{({\text{Se}}_{1\ensuremath{-}x}{\text{Te}}_{x})}_{0.82}$},}\
  }\href {\doibase 10.1103/PhysRevB.78.224503} {\bibfield  {journal} {\bibinfo
  {journal} {Phys. Rev. B}\ }\textbf {\bibinfo {volume} {78}},\ \bibinfo
  {pages} {224503} (\bibinfo {year} {2008})}\BibitemShut {NoStop}%
\bibitem [{\citenamefont {Dai}(2015)}]{Dai2015}%
  \BibitemOpen
  \bibfield  {author} {\bibinfo {author} {\bibfnamefont {Pengcheng}\
  \bibnamefont {Dai}},\ }\bibfield  {title} {\enquote {\bibinfo {title}
  {Antiferromagnetic order and spin dynamics in iron-based superconductors},}\
  }\href {\doibase 10.1103/RevModPhys.87.855} {\bibfield  {journal} {\bibinfo
  {journal} {Rev. Mod. Phys.}\ }\textbf {\bibinfo {volume} {87}},\ \bibinfo
  {pages} {855--896} (\bibinfo {year} {2015})}\BibitemShut {NoStop}%
\bibitem [{\citenamefont {McQueen}\ \emph {et~al.}(2009)\citenamefont
  {McQueen}, \citenamefont {Williams}, \citenamefont {Stephens}, \citenamefont
  {Tao}, \citenamefont {Zhu}, \citenamefont {Ksenofontov}, \citenamefont
  {Casper}, \citenamefont {Felser},\ and\ \citenamefont {Cava}}]{McQueen2009}%
  \BibitemOpen
  \bibfield  {author} {\bibinfo {author} {\bibfnamefont {T.~M.}\ \bibnamefont
  {McQueen}}, \bibinfo {author} {\bibfnamefont {A.~J.}\ \bibnamefont
  {Williams}}, \bibinfo {author} {\bibfnamefont {P.~W.}\ \bibnamefont
  {Stephens}}, \bibinfo {author} {\bibfnamefont {J.}~\bibnamefont {Tao}},
  \bibinfo {author} {\bibfnamefont {Y.}~\bibnamefont {Zhu}}, \bibinfo {author}
  {\bibfnamefont {V.}~\bibnamefont {Ksenofontov}}, \bibinfo {author}
  {\bibfnamefont {F.}~\bibnamefont {Casper}}, \bibinfo {author} {\bibfnamefont
  {C.}~\bibnamefont {Felser}}, \ and\ \bibinfo {author} {\bibfnamefont {R.~J.}\
  \bibnamefont {Cava}},\ }\bibfield  {title} {\enquote {\bibinfo {title}
  {Tetragonal-to-orthorhombic structural phase transition at 90 k in the
  superconductor ${\mathrm{fe}}_{1.01}\mathrm{Se}$},}\ }\href {\doibase
  10.1103/PhysRevLett.103.057002} {\bibfield  {journal} {\bibinfo  {journal}
  {Phys. Rev. Lett.}\ }\textbf {\bibinfo {volume} {103}},\ \bibinfo {pages}
  {057002} (\bibinfo {year} {2009})}\BibitemShut {NoStop}%
\bibitem [{\citenamefont {Medvedev}\ \emph {et~al.}(2009)\citenamefont
  {Medvedev}, \citenamefont {McQueen}, \citenamefont {Troyan}, \citenamefont
  {Palasyuk}, \citenamefont {Eremets}, \citenamefont {Cava}, \citenamefont
  {Naghavi}, \citenamefont {Casper}, \citenamefont {Ksenofontov}, \citenamefont
  {Wortmann},\ and\ \citenamefont {Felser}}]{Medvedev2009}%
  \BibitemOpen
  \bibfield  {author} {\bibinfo {author} {\bibfnamefont {S.}~\bibnamefont
  {Medvedev}}, \bibinfo {author} {\bibfnamefont {T.~M.}\ \bibnamefont
  {McQueen}}, \bibinfo {author} {\bibfnamefont {I.~A.}\ \bibnamefont {Troyan}},
  \bibinfo {author} {\bibfnamefont {T.}~\bibnamefont {Palasyuk}}, \bibinfo
  {author} {\bibfnamefont {M.~I.}\ \bibnamefont {Eremets}}, \bibinfo {author}
  {\bibfnamefont {R.~J.}\ \bibnamefont {Cava}}, \bibinfo {author}
  {\bibfnamefont {S.}~\bibnamefont {Naghavi}}, \bibinfo {author} {\bibfnamefont
  {F.}~\bibnamefont {Casper}}, \bibinfo {author} {\bibfnamefont
  {V.}~\bibnamefont {Ksenofontov}}, \bibinfo {author} {\bibfnamefont
  {G.}~\bibnamefont {Wortmann}}, \ and\ \bibinfo {author} {\bibfnamefont
  {C.}~\bibnamefont {Felser}},\ }\bibfield  {title} {\enquote {\bibinfo {title}
  {Electronic and magnetic phase diagram of $\beta$-fe1.01se with
  superconductivity at 36.7 k under pressure},}\ }\href {\doibase
  10.1038/nmat2491} {\bibfield  {journal} {\bibinfo  {journal} {Nature
  Materials}\ }\textbf {\bibinfo {volume} {8}},\ \bibinfo {pages} {630--633}
  (\bibinfo {year} {2009})}\BibitemShut {NoStop}%
\bibitem [{\citenamefont {B\"ohmer}\ \emph {et~al.}(2015)\citenamefont
  {B\"ohmer}, \citenamefont {Arai}, \citenamefont {Hardy}, \citenamefont
  {Hattori}, \citenamefont {Iye}, \citenamefont {Wolf}, \citenamefont
  {L\"ohneysen}, \citenamefont {Ishida},\ and\ \citenamefont
  {Meingast}}]{Bohmer2015}%
  \BibitemOpen
  \bibfield  {author} {\bibinfo {author} {\bibfnamefont {A.~E.}\ \bibnamefont
  {B\"ohmer}}, \bibinfo {author} {\bibfnamefont {T.}~\bibnamefont {Arai}},
  \bibinfo {author} {\bibfnamefont {F.}~\bibnamefont {Hardy}}, \bibinfo
  {author} {\bibfnamefont {T.}~\bibnamefont {Hattori}}, \bibinfo {author}
  {\bibfnamefont {T.}~\bibnamefont {Iye}}, \bibinfo {author} {\bibfnamefont
  {T.}~\bibnamefont {Wolf}}, \bibinfo {author} {\bibfnamefont {H.~v.}\
  \bibnamefont {L\"ohneysen}}, \bibinfo {author} {\bibfnamefont
  {K.}~\bibnamefont {Ishida}}, \ and\ \bibinfo {author} {\bibfnamefont
  {C.}~\bibnamefont {Meingast}},\ }\bibfield  {title} {\enquote {\bibinfo
  {title} {Origin of the tetragonal-to-orthorhombic phase transition in fese: A
  combined thermodynamic and nmr study of nematicity},}\ }\href {\doibase
  10.1103/PhysRevLett.114.027001} {\bibfield  {journal} {\bibinfo  {journal}
  {Phys. Rev. Lett.}\ }\textbf {\bibinfo {volume} {114}},\ \bibinfo {pages}
  {027001} (\bibinfo {year} {2015})}\BibitemShut {NoStop}%
\bibitem [{\citenamefont {Baek}\ \emph {et~al.}(2015)\citenamefont {Baek},
  \citenamefont {Efremov}, \citenamefont {Ok}, \citenamefont {Kim},
  \citenamefont {Van Den~Brink},\ and\ \citenamefont {B{\"u}chner}}]{Baek2015}%
  \BibitemOpen
  \bibfield  {author} {\bibinfo {author} {\bibfnamefont {SH}~\bibnamefont
  {Baek}}, \bibinfo {author} {\bibfnamefont {DV}~\bibnamefont {Efremov}},
  \bibinfo {author} {\bibfnamefont {JM}~\bibnamefont {Ok}}, \bibinfo {author}
  {\bibfnamefont {JS}~\bibnamefont {Kim}}, \bibinfo {author} {\bibfnamefont
  {Jeroen}\ \bibnamefont {Van Den~Brink}}, \ and\ \bibinfo {author}
  {\bibfnamefont {B}~\bibnamefont {B{\"u}chner}},\ }\bibfield  {title}
  {\enquote {\bibinfo {title} {Orbital-driven nematicity in fese},}\ }\href
  {https://doi.org/10.1038/nmat4138} {\bibfield  {journal} {\bibinfo  {journal}
  {Nature materials}\ }\textbf {\bibinfo {volume} {14}},\ \bibinfo {pages}
  {210} (\bibinfo {year} {2015})}\BibitemShut {NoStop}%
\bibitem [{\citenamefont {Nakayama}\ \emph {et~al.}(2014)\citenamefont
  {Nakayama}, \citenamefont {Miyata}, \citenamefont {Phan}, \citenamefont
  {Sato}, \citenamefont {Tanabe}, \citenamefont {Urata}, \citenamefont
  {Tanigaki},\ and\ \citenamefont {Takahashi}}]{Nakayama2014}%
  \BibitemOpen
  \bibfield  {author} {\bibinfo {author} {\bibfnamefont {K.}~\bibnamefont
  {Nakayama}}, \bibinfo {author} {\bibfnamefont {Y.}~\bibnamefont {Miyata}},
  \bibinfo {author} {\bibfnamefont {G.~N.}\ \bibnamefont {Phan}}, \bibinfo
  {author} {\bibfnamefont {T.}~\bibnamefont {Sato}}, \bibinfo {author}
  {\bibfnamefont {Y.}~\bibnamefont {Tanabe}}, \bibinfo {author} {\bibfnamefont
  {T.}~\bibnamefont {Urata}}, \bibinfo {author} {\bibfnamefont
  {K.}~\bibnamefont {Tanigaki}}, \ and\ \bibinfo {author} {\bibfnamefont
  {T.}~\bibnamefont {Takahashi}},\ }\bibfield  {title} {\enquote {\bibinfo
  {title} {Reconstruction of band structure induced by electronic nematicity in
  an fese superconductor},}\ }\href {\doibase 10.1103/PhysRevLett.113.237001}
  {\bibfield  {journal} {\bibinfo  {journal} {Phys. Rev. Lett.}\ }\textbf
  {\bibinfo {volume} {113}},\ \bibinfo {pages} {237001} (\bibinfo {year}
  {2014})}\BibitemShut {NoStop}%
\bibitem [{\citenamefont {Shimojima}\ \emph {et~al.}(2014)\citenamefont
  {Shimojima}, \citenamefont {Suzuki}, \citenamefont {Sonobe}, \citenamefont
  {Nakamura}, \citenamefont {Sakano}, \citenamefont {Omachi}, \citenamefont
  {Yoshioka}, \citenamefont {Kuwata-Gonokami}, \citenamefont {Ono},
  \citenamefont {Kumigashira}, \citenamefont {B\"ohmer}, \citenamefont {Hardy},
  \citenamefont {Wolf}, \citenamefont {Meingast}, \citenamefont {L\"ohneysen},
  \citenamefont {Ikeda},\ and\ \citenamefont {Ishizaka}}]{Shimojima2014}%
  \BibitemOpen
  \bibfield  {author} {\bibinfo {author} {\bibfnamefont {T.}~\bibnamefont
  {Shimojima}}, \bibinfo {author} {\bibfnamefont {Y.}~\bibnamefont {Suzuki}},
  \bibinfo {author} {\bibfnamefont {T.}~\bibnamefont {Sonobe}}, \bibinfo
  {author} {\bibfnamefont {A.}~\bibnamefont {Nakamura}}, \bibinfo {author}
  {\bibfnamefont {M.}~\bibnamefont {Sakano}}, \bibinfo {author} {\bibfnamefont
  {J.}~\bibnamefont {Omachi}}, \bibinfo {author} {\bibfnamefont
  {K.}~\bibnamefont {Yoshioka}}, \bibinfo {author} {\bibfnamefont
  {M.}~\bibnamefont {Kuwata-Gonokami}}, \bibinfo {author} {\bibfnamefont
  {K.}~\bibnamefont {Ono}}, \bibinfo {author} {\bibfnamefont {H.}~\bibnamefont
  {Kumigashira}}, \bibinfo {author} {\bibfnamefont {A.~E.}\ \bibnamefont
  {B\"ohmer}}, \bibinfo {author} {\bibfnamefont {F.}~\bibnamefont {Hardy}},
  \bibinfo {author} {\bibfnamefont {T.}~\bibnamefont {Wolf}}, \bibinfo {author}
  {\bibfnamefont {C.}~\bibnamefont {Meingast}}, \bibinfo {author}
  {\bibfnamefont {H.~v.}\ \bibnamefont {L\"ohneysen}}, \bibinfo {author}
  {\bibfnamefont {H.}~\bibnamefont {Ikeda}}, \ and\ \bibinfo {author}
  {\bibfnamefont {K.}~\bibnamefont {Ishizaka}},\ }\bibfield  {title} {\enquote
  {\bibinfo {title} {Lifting of xz/yz orbital degeneracy at the structural
  transition in detwinned fese},}\ }\href {\doibase 10.1103/PhysRevB.90.121111}
  {\bibfield  {journal} {\bibinfo  {journal} {Phys. Rev. B}\ }\textbf {\bibinfo
  {volume} {90}},\ \bibinfo {pages} {121111} (\bibinfo {year}
  {2014})}\BibitemShut {NoStop}%
\bibitem [{\citenamefont {Watson}\ \emph {et~al.}(2015)\citenamefont {Watson},
  \citenamefont {Kim}, \citenamefont {Haghighirad}, \citenamefont {Davies},
  \citenamefont {McCollam}, \citenamefont {Narayanan}, \citenamefont {Blake},
  \citenamefont {Chen}, \citenamefont {Ghannadzadeh}, \citenamefont
  {Schofield}, \citenamefont {Hoesch}, \citenamefont {Meingast}, \citenamefont
  {Wolf},\ and\ \citenamefont {Coldea}}]{Watson2015a}%
  \BibitemOpen
  \bibfield  {author} {\bibinfo {author} {\bibfnamefont {M.~D.}\ \bibnamefont
  {Watson}}, \bibinfo {author} {\bibfnamefont {T.~K.}\ \bibnamefont {Kim}},
  \bibinfo {author} {\bibfnamefont {A.~A.}\ \bibnamefont {Haghighirad}},
  \bibinfo {author} {\bibfnamefont {N.~R.}\ \bibnamefont {Davies}}, \bibinfo
  {author} {\bibfnamefont {A.}~\bibnamefont {McCollam}}, \bibinfo {author}
  {\bibfnamefont {A.}~\bibnamefont {Narayanan}}, \bibinfo {author}
  {\bibfnamefont {S.~F.}\ \bibnamefont {Blake}}, \bibinfo {author}
  {\bibfnamefont {Y.~L.}\ \bibnamefont {Chen}}, \bibinfo {author}
  {\bibfnamefont {S.}~\bibnamefont {Ghannadzadeh}}, \bibinfo {author}
  {\bibfnamefont {A.~J.}\ \bibnamefont {Schofield}}, \bibinfo {author}
  {\bibfnamefont {M.}~\bibnamefont {Hoesch}}, \bibinfo {author} {\bibfnamefont
  {C.}~\bibnamefont {Meingast}}, \bibinfo {author} {\bibfnamefont
  {T.}~\bibnamefont {Wolf}}, \ and\ \bibinfo {author} {\bibfnamefont {A.~I.}\
  \bibnamefont {Coldea}},\ }\bibfield  {title} {\enquote {\bibinfo {title}
  {Emergence of the nematic electronic state in fese},}\ }\href {\doibase
  10.1103/PhysRevB.91.155106} {\bibfield  {journal} {\bibinfo  {journal} {Phys.
  Rev. B}\ }\textbf {\bibinfo {volume} {91}},\ \bibinfo {pages} {155106}
  (\bibinfo {year} {2015})}\BibitemShut {NoStop}%
\bibitem [{\citenamefont {Terashima}\ \emph {et~al.}(2015)\citenamefont
  {Terashima}, \citenamefont {Kikugawa}, \citenamefont {Kasahara},
  \citenamefont {Watashige}, \citenamefont {Shibauchi}, \citenamefont
  {Matsuda}, \citenamefont {Wolf}, \citenamefont {B\"ohmer}, \citenamefont
  {Hardy}, \citenamefont {Meingast}, \citenamefont {L\"ohneysen},\ and\
  \citenamefont {Uji}}]{Terashima2015}%
  \BibitemOpen
  \bibfield  {author} {\bibinfo {author} {\bibfnamefont {Taichi}\ \bibnamefont
  {Terashima}}, \bibinfo {author} {\bibfnamefont {Naoki}\ \bibnamefont
  {Kikugawa}}, \bibinfo {author} {\bibfnamefont {Shigeru}\ \bibnamefont
  {Kasahara}}, \bibinfo {author} {\bibfnamefont {Tatsuya}\ \bibnamefont
  {Watashige}}, \bibinfo {author} {\bibfnamefont {Takasada}\ \bibnamefont
  {Shibauchi}}, \bibinfo {author} {\bibfnamefont {Yuji}\ \bibnamefont
  {Matsuda}}, \bibinfo {author} {\bibfnamefont {Thomas}\ \bibnamefont {Wolf}},
  \bibinfo {author} {\bibfnamefont {Anna~E.}\ \bibnamefont {B\"ohmer}},
  \bibinfo {author} {\bibfnamefont {Fr\'{e}d\'{e}ric}\ \bibnamefont {Hardy}},
  \bibinfo {author} {\bibfnamefont {Christoph}\ \bibnamefont {Meingast}},
  \bibinfo {author} {\bibfnamefont {Hilbert~v.}\ \bibnamefont {L\"ohneysen}}, \
  and\ \bibinfo {author} {\bibfnamefont {Shinya}\ \bibnamefont {Uji}},\
  }\bibfield  {title} {\enquote {\bibinfo {title} {Pressure-induced
  antiferromagnetic transition and phase diagram in fese},}\ }\href {\doibase
  10.7566/JPSJ.84.063701} {\bibfield  {journal} {\bibinfo  {journal} {Journal
  of the Physical Society of Japan}\ }\textbf {\bibinfo {volume} {84}},\
  \bibinfo {pages} {063701} (\bibinfo {year} {2015})}\BibitemShut {NoStop}%
\bibitem [{\citenamefont {B{\"o}hmer}\ and\ \citenamefont
  {Kreisel}(2017)}]{bohmer2017}%
  \BibitemOpen
  \bibfield  {author} {\bibinfo {author} {\bibfnamefont {Anna~E}\ \bibnamefont
  {B{\"o}hmer}}\ and\ \bibinfo {author} {\bibfnamefont {Andreas}\ \bibnamefont
  {Kreisel}},\ }\bibfield  {title} {\enquote {\bibinfo {title} {Nematicity,
  magnetism and superconductivity in fese},}\ }\href@noop {} {\bibfield
  {journal} {\bibinfo  {journal} {Journal of Physics: Condensed Matter}\
  }\textbf {\bibinfo {volume} {30}},\ \bibinfo {pages} {023001} (\bibinfo
  {year} {2017})}\BibitemShut {NoStop}%
\bibitem [{\citenamefont {Coldea}\ and\ \citenamefont
  {Watson}(2018)}]{coldea2018}%
  \BibitemOpen
  \bibfield  {author} {\bibinfo {author} {\bibfnamefont {Amalia~I}\
  \bibnamefont {Coldea}}\ and\ \bibinfo {author} {\bibfnamefont {Matthew~D}\
  \bibnamefont {Watson}},\ }\bibfield  {title} {\enquote {\bibinfo {title} {The
  key ingredients of the electronic structure of fese},}\ }\href@noop {}
  {\bibfield  {journal} {\bibinfo  {journal} {Annual Review of Condensed Matter
  Physics}\ }\textbf {\bibinfo {volume} {9}},\ \bibinfo {pages} {125--146}
  (\bibinfo {year} {2018})}\BibitemShut {NoStop}%
\bibitem [{\citenamefont {Chen}\ \emph {et~al.}(2019)\citenamefont {Chen},
  \citenamefont {Chen}, \citenamefont {Kreisel}, \citenamefont {Lu},
  \citenamefont {Schneidewind}, \citenamefont {Qiu}, \citenamefont {Park},
  \citenamefont {Perring}, \citenamefont {Stewart}, \citenamefont {Cao},
  \citenamefont {Zhang}, \citenamefont {Li}, \citenamefont {Rong},
  \citenamefont {Wei}, \citenamefont {Andersen}, \citenamefont {Hirschfeld},
  \citenamefont {Broholm},\ and\ \citenamefont {Dai}}]{chen2019}%
  \BibitemOpen
  \bibfield  {author} {\bibinfo {author} {\bibfnamefont {Tong}\ \bibnamefont
  {Chen}}, \bibinfo {author} {\bibfnamefont {Youzhe}\ \bibnamefont {Chen}},
  \bibinfo {author} {\bibfnamefont {Andreas}\ \bibnamefont {Kreisel}}, \bibinfo
  {author} {\bibfnamefont {Xingye}\ \bibnamefont {Lu}}, \bibinfo {author}
  {\bibfnamefont {Astrid}\ \bibnamefont {Schneidewind}}, \bibinfo {author}
  {\bibfnamefont {Yiming}\ \bibnamefont {Qiu}}, \bibinfo {author}
  {\bibfnamefont {J.~T.}\ \bibnamefont {Park}}, \bibinfo {author}
  {\bibfnamefont {Toby~G.}\ \bibnamefont {Perring}}, \bibinfo {author}
  {\bibfnamefont {J.~Ross}\ \bibnamefont {Stewart}}, \bibinfo {author}
  {\bibfnamefont {Huibo}\ \bibnamefont {Cao}}, \bibinfo {author} {\bibfnamefont
  {Rui}\ \bibnamefont {Zhang}}, \bibinfo {author} {\bibfnamefont
  {Yu}~\bibnamefont {Li}}, \bibinfo {author} {\bibfnamefont {Yan}\ \bibnamefont
  {Rong}}, \bibinfo {author} {\bibfnamefont {Yuan}\ \bibnamefont {Wei}},
  \bibinfo {author} {\bibfnamefont {Brian~M.}\ \bibnamefont {Andersen}},
  \bibinfo {author} {\bibfnamefont {P.~J.}\ \bibnamefont {Hirschfeld}},
  \bibinfo {author} {\bibfnamefont {Collin}\ \bibnamefont {Broholm}}, \ and\
  \bibinfo {author} {\bibfnamefont {Pengcheng}\ \bibnamefont {Dai}},\
  }\bibfield  {title} {\enquote {\bibinfo {title} {Anisotropic spin
  fluctuations in detwinned fese},}\ }\href {\doibase
  10.1038/s41563-019-0369-5} {\bibfield  {journal} {\bibinfo  {journal} {Nature
  Materials}\ }\textbf {\bibinfo {volume} {18}},\ \bibinfo {pages} {709--716}
  (\bibinfo {year} {2019})}\BibitemShut {NoStop}%
\bibitem [{\citenamefont {Yi}\ \emph {et~al.}(2019)\citenamefont {Yi},
  \citenamefont {Pfau}, \citenamefont {Zhang}, \citenamefont {He},
  \citenamefont {Wu}, \citenamefont {Chen}, \citenamefont {Ye}, \citenamefont
  {Hashimoto}, \citenamefont {Yu}, \citenamefont {Si}, \citenamefont {Lee},
  \citenamefont {Dai}, \citenamefont {Shen}, \citenamefont {Lu},\ and\
  \citenamefont {Birgeneau}}]{Yi2019}%
  \BibitemOpen
  \bibfield  {author} {\bibinfo {author} {\bibfnamefont {M.}~\bibnamefont
  {Yi}}, \bibinfo {author} {\bibfnamefont {H.}~\bibnamefont {Pfau}}, \bibinfo
  {author} {\bibfnamefont {Y.}~\bibnamefont {Zhang}}, \bibinfo {author}
  {\bibfnamefont {Y.}~\bibnamefont {He}}, \bibinfo {author} {\bibfnamefont
  {H.}~\bibnamefont {Wu}}, \bibinfo {author} {\bibfnamefont {T.}~\bibnamefont
  {Chen}}, \bibinfo {author} {\bibfnamefont {Z.~R.}\ \bibnamefont {Ye}},
  \bibinfo {author} {\bibfnamefont {M.}~\bibnamefont {Hashimoto}}, \bibinfo
  {author} {\bibfnamefont {R.}~\bibnamefont {Yu}}, \bibinfo {author}
  {\bibfnamefont {Q.}~\bibnamefont {Si}}, \bibinfo {author} {\bibfnamefont
  {D.-H.}\ \bibnamefont {Lee}}, \bibinfo {author} {\bibfnamefont {Pengcheng}\
  \bibnamefont {Dai}}, \bibinfo {author} {\bibfnamefont {Z.-X.}\ \bibnamefont
  {Shen}}, \bibinfo {author} {\bibfnamefont {D.~H.}\ \bibnamefont {Lu}}, \ and\
  \bibinfo {author} {\bibfnamefont {R.~J.}\ \bibnamefont {Birgeneau}},\
  }\bibfield  {title} {\enquote {\bibinfo {title} {Nematic energy scale and the
  missing electron pocket in fese},}\ }\href {\doibase
  10.1103/PhysRevX.9.041049} {\bibfield  {journal} {\bibinfo  {journal} {Phys.
  Rev. X}\ }\textbf {\bibinfo {volume} {9}},\ \bibinfo {pages} {041049}
  (\bibinfo {year} {2019})}\BibitemShut {NoStop}%
\bibitem [{\citenamefont {Yu}\ and\ \citenamefont {Si}(2015)}]{YuSi_AFQ}%
  \BibitemOpen
  \bibfield  {author} {\bibinfo {author} {\bibfnamefont {Rong}\ \bibnamefont
  {Yu}}\ and\ \bibinfo {author} {\bibfnamefont {Qimiao}\ \bibnamefont {Si}},\
  }\bibfield  {title} {\enquote {\bibinfo {title} {Antiferroquadrupolar and
  ising-nematic orders of a frustrated bilinear-biquadratic heisenberg model
  and implications for the magnetism of fese},}\ }\href {\doibase
  10.1103/PhysRevLett.115.116401} {\bibfield  {journal} {\bibinfo  {journal}
  {Phys. Rev. Lett.}\ }\textbf {\bibinfo {volume} {115}},\ \bibinfo {pages}
  {116401} (\bibinfo {year} {2015})}\BibitemShut {NoStop}%
\bibitem [{\citenamefont {Wang}\ \emph {et~al.}(2015)\citenamefont {Wang},
  \citenamefont {Kivelson},\ and\ \citenamefont {Lee}}]{FaWang2015}%
  \BibitemOpen
  \bibfield  {author} {\bibinfo {author} {\bibfnamefont {Fa}~\bibnamefont
  {Wang}}, \bibinfo {author} {\bibfnamefont {Steven~A}\ \bibnamefont
  {Kivelson}}, \ and\ \bibinfo {author} {\bibfnamefont {Dung-Hai}\ \bibnamefont
  {Lee}},\ }\bibfield  {title} {\enquote {\bibinfo {title} {Nematicity and
  quantum paramagnetism in fese},}\ }\href {https://doi.org/10.1038/nphys3456}
  {\bibfield  {journal} {\bibinfo  {journal} {Nature Physics}\ }\textbf
  {\bibinfo {volume} {11}},\ \bibinfo {pages} {959} (\bibinfo {year}
  {2015})}\BibitemShut {NoStop}%
\bibitem [{\citenamefont {Glasbrenner}\ \emph {et~al.}(2015)\citenamefont
  {Glasbrenner}, \citenamefont {Mazin}, \citenamefont {Jeschke}, \citenamefont
  {Hirschfeld}, \citenamefont {Fernandes},\ and\ \citenamefont
  {Valent{\'\i}}}]{Glasbrenner2015}%
  \BibitemOpen
  \bibfield  {author} {\bibinfo {author} {\bibfnamefont {JK}~\bibnamefont
  {Glasbrenner}}, \bibinfo {author} {\bibfnamefont {II}~\bibnamefont {Mazin}},
  \bibinfo {author} {\bibfnamefont {Harald~O}\ \bibnamefont {Jeschke}},
  \bibinfo {author} {\bibfnamefont {PJ}~\bibnamefont {Hirschfeld}}, \bibinfo
  {author} {\bibfnamefont {RM}~\bibnamefont {Fernandes}}, \ and\ \bibinfo
  {author} {\bibfnamefont {Roser}\ \bibnamefont {Valent{\'\i}}},\ }\bibfield
  {title} {\enquote {\bibinfo {title} {Effect of magnetic frustration on
  nematicity and superconductivity in iron chalcogenides},}\ }\href
  {https://doi.org/10.1038/nphys3434} {\bibfield  {journal} {\bibinfo
  {journal} {Nature Physics}\ }\textbf {\bibinfo {volume} {11}},\ \bibinfo
  {pages} {953} (\bibinfo {year} {2015})}\BibitemShut {NoStop}%
\bibitem [{\citenamefont {Rahn}\ \emph {et~al.}(2015)\citenamefont {Rahn},
  \citenamefont {Ewings}, \citenamefont {Sedlmaier}, \citenamefont {Clarke},\
  and\ \citenamefont {Boothroyd}}]{Rahn2015}%
  \BibitemOpen
  \bibfield  {author} {\bibinfo {author} {\bibfnamefont {M.~C.}\ \bibnamefont
  {Rahn}}, \bibinfo {author} {\bibfnamefont {R.~A.}\ \bibnamefont {Ewings}},
  \bibinfo {author} {\bibfnamefont {S.~J.}\ \bibnamefont {Sedlmaier}}, \bibinfo
  {author} {\bibfnamefont {S.~J.}\ \bibnamefont {Clarke}}, \ and\ \bibinfo
  {author} {\bibfnamefont {A.~T.}\ \bibnamefont {Boothroyd}},\ }\bibfield
  {title} {\enquote {\bibinfo {title} {Strong $(\ensuremath{\pi},0)$ spin
  fluctuations in $\ensuremath{\beta}\ensuremath{-}\mathrm{FeSe}$ observed by
  neutron spectroscopy},}\ }\href {\doibase 10.1103/PhysRevB.91.180501}
  {\bibfield  {journal} {\bibinfo  {journal} {Phys. Rev. B}\ }\textbf {\bibinfo
  {volume} {91}},\ \bibinfo {pages} {180501} (\bibinfo {year}
  {2015})}\BibitemShut {NoStop}%
\bibitem [{\citenamefont {Wang}\ \emph
  {et~al.}(2016{\natexlab{a}})\citenamefont {Wang}, \citenamefont {Shen},
  \citenamefont {Pan}, \citenamefont {Hao}, \citenamefont {Ma}, \citenamefont
  {Zhou}, \citenamefont {Steffens}, \citenamefont {Schmalzl}, \citenamefont
  {Forrest}, \citenamefont {Abdel-Hafiez}, \citenamefont {Chen}, \citenamefont
  {Chareev}, \citenamefont {Vasiliev}, \citenamefont {Bourges}, \citenamefont
  {Sidis}, \citenamefont {Cao},\ and\ \citenamefont {Zhao}}]{WangZhao2016}%
  \BibitemOpen
  \bibfield  {author} {\bibinfo {author} {\bibfnamefont {Qisi}\ \bibnamefont
  {Wang}}, \bibinfo {author} {\bibfnamefont {Yao}\ \bibnamefont {Shen}},
  \bibinfo {author} {\bibfnamefont {Bingying}\ \bibnamefont {Pan}}, \bibinfo
  {author} {\bibfnamefont {Yiqing}\ \bibnamefont {Hao}}, \bibinfo {author}
  {\bibfnamefont {Mingwei}\ \bibnamefont {Ma}}, \bibinfo {author}
  {\bibfnamefont {Fang}\ \bibnamefont {Zhou}}, \bibinfo {author} {\bibfnamefont
  {P.}~\bibnamefont {Steffens}}, \bibinfo {author} {\bibfnamefont
  {K.}~\bibnamefont {Schmalzl}}, \bibinfo {author} {\bibfnamefont {T.~R.}\
  \bibnamefont {Forrest}}, \bibinfo {author} {\bibfnamefont {M.}~\bibnamefont
  {Abdel-Hafiez}}, \bibinfo {author} {\bibfnamefont {Xiaojia}\ \bibnamefont
  {Chen}}, \bibinfo {author} {\bibfnamefont {D.~A.}\ \bibnamefont {Chareev}},
  \bibinfo {author} {\bibfnamefont {A.~N.}\ \bibnamefont {Vasiliev}}, \bibinfo
  {author} {\bibfnamefont {P.}~\bibnamefont {Bourges}}, \bibinfo {author}
  {\bibfnamefont {Y.}~\bibnamefont {Sidis}}, \bibinfo {author} {\bibfnamefont
  {Huibo}\ \bibnamefont {Cao}}, \ and\ \bibinfo {author} {\bibfnamefont {Jun}\
  \bibnamefont {Zhao}},\ }\bibfield  {title} {\enquote {\bibinfo {title}
  {Strong interplay between stripe spin fluctuations, nematicity and
  superconductivity in fese},}\ }\href {\doibase 10.1038/nmat4492} {\bibfield
  {journal} {\bibinfo  {journal} {Nature Materials}\ }\textbf {\bibinfo
  {volume} {15}},\ \bibinfo {pages} {159--163} (\bibinfo {year}
  {2016}{\natexlab{a}})}\BibitemShut {NoStop}%
\bibitem [{\citenamefont {Wang}\ \emph
  {et~al.}(2016{\natexlab{b}})\citenamefont {Wang}, \citenamefont {Shen},
  \citenamefont {Pan}, \citenamefont {Zhang}, \citenamefont {Ikeuchi},
  \citenamefont {Iida}, \citenamefont {Christianson}, \citenamefont {Walker},
  \citenamefont {Adroja}, \citenamefont {Abdel-Hafiez}, \citenamefont {Chen},
  \citenamefont {Chareev}, \citenamefont {Vasiliev},\ and\ \citenamefont
  {Zhao}}]{QWang2015b}%
  \BibitemOpen
  \bibfield  {author} {\bibinfo {author} {\bibfnamefont {Qisi}\ \bibnamefont
  {Wang}}, \bibinfo {author} {\bibfnamefont {Yao}\ \bibnamefont {Shen}},
  \bibinfo {author} {\bibfnamefont {Bingying}\ \bibnamefont {Pan}}, \bibinfo
  {author} {\bibfnamefont {Xiaowen}\ \bibnamefont {Zhang}}, \bibinfo {author}
  {\bibfnamefont {K.}~\bibnamefont {Ikeuchi}}, \bibinfo {author} {\bibfnamefont
  {K.}~\bibnamefont {Iida}}, \bibinfo {author} {\bibfnamefont {A.~D.}\
  \bibnamefont {Christianson}}, \bibinfo {author} {\bibfnamefont {H.~C.}\
  \bibnamefont {Walker}}, \bibinfo {author} {\bibfnamefont {D.~T.}\
  \bibnamefont {Adroja}}, \bibinfo {author} {\bibfnamefont {M.}~\bibnamefont
  {Abdel-Hafiez}}, \bibinfo {author} {\bibfnamefont {Xiaojia}\ \bibnamefont
  {Chen}}, \bibinfo {author} {\bibfnamefont {D.~A.}\ \bibnamefont {Chareev}},
  \bibinfo {author} {\bibfnamefont {A.~N.}\ \bibnamefont {Vasiliev}}, \ and\
  \bibinfo {author} {\bibfnamefont {Jun}\ \bibnamefont {Zhao}},\ }\bibfield
  {title} {\enquote {\bibinfo {title} {Magnetic ground state of fese},}\ }\href
  {\doibase 10.1038/ncomms12182} {\bibfield  {journal} {\bibinfo  {journal}
  {Nature Communications}\ }\textbf {\bibinfo {volume} {7}},\ \bibinfo {pages}
  {12182} (\bibinfo {year} {2016}{\natexlab{b}})}\BibitemShut {NoStop}%
\bibitem [{\citenamefont {Mukherjee}\ \emph {et~al.}(2015)\citenamefont
  {Mukherjee}, \citenamefont {Kreisel}, \citenamefont {Hirschfeld},\ and\
  \citenamefont {Andersen}}]{Mukherjee2015}%
  \BibitemOpen
  \bibfield  {author} {\bibinfo {author} {\bibfnamefont {Shantanu}\
  \bibnamefont {Mukherjee}}, \bibinfo {author} {\bibfnamefont {A.}~\bibnamefont
  {Kreisel}}, \bibinfo {author} {\bibfnamefont {P.~J.}\ \bibnamefont
  {Hirschfeld}}, \ and\ \bibinfo {author} {\bibfnamefont {Brian~M.}\
  \bibnamefont {Andersen}},\ }\bibfield  {title} {\enquote {\bibinfo {title}
  {Model of electronic structure and superconductivity in orbitally ordered
  fese},}\ }\href {\doibase 10.1103/PhysRevLett.115.026402} {\bibfield
  {journal} {\bibinfo  {journal} {Phys. Rev. Lett.}\ }\textbf {\bibinfo
  {volume} {115}},\ \bibinfo {pages} {026402} (\bibinfo {year}
  {2015})}\BibitemShut {NoStop}%
\bibitem [{\citenamefont {Yu}\ \emph {et~al.}(2012)\citenamefont {Yu},
  \citenamefont {Wang}, \citenamefont {Goswami}, \citenamefont {Nevidomskyy},
  \citenamefont {Si},\ and\ \citenamefont {Abrahams}}]{Yu2012}%
  \BibitemOpen
  \bibfield  {author} {\bibinfo {author} {\bibfnamefont {Rong}\ \bibnamefont
  {Yu}}, \bibinfo {author} {\bibfnamefont {Zhentao}\ \bibnamefont {Wang}},
  \bibinfo {author} {\bibfnamefont {Pallab}\ \bibnamefont {Goswami}}, \bibinfo
  {author} {\bibfnamefont {Andriy~H.}\ \bibnamefont {Nevidomskyy}}, \bibinfo
  {author} {\bibfnamefont {Qimiao}\ \bibnamefont {Si}}, \ and\ \bibinfo
  {author} {\bibfnamefont {Elihu}\ \bibnamefont {Abrahams}},\ }\bibfield
  {title} {\enquote {\bibinfo {title} {Spin dynamics of a
  ${J}_{1}$-${J}_{2}$-$k$ model for the paramagnetic phase of iron
  pnictides},}\ }\href {\doibase 10.1103/PhysRevB.86.085148} {\bibfield
  {journal} {\bibinfo  {journal} {Phys. Rev. B}\ }\textbf {\bibinfo {volume}
  {86}},\ \bibinfo {pages} {085148} (\bibinfo {year} {2012})}\BibitemShut
  {NoStop}%
\bibitem [{\citenamefont {Wysocki}\ \emph {et~al.}(2011)\citenamefont
  {Wysocki}, \citenamefont {Belashchenko},\ and\ \citenamefont
  {Antropov}}]{Wysocki2011}%
  \BibitemOpen
  \bibfield  {author} {\bibinfo {author} {\bibfnamefont {Aleksander~L}\
  \bibnamefont {Wysocki}}, \bibinfo {author} {\bibfnamefont {Kirill~D}\
  \bibnamefont {Belashchenko}}, \ and\ \bibinfo {author} {\bibfnamefont
  {Vladimir~P}\ \bibnamefont {Antropov}},\ }\bibfield  {title} {\enquote
  {\bibinfo {title} {Consistent model of magnetism in ferropnictides},}\ }\href
  {https://doi.org/10.1038/nphys1933} {\bibfield  {journal} {\bibinfo
  {journal} {Nature Physics}\ }\textbf {\bibinfo {volume} {7}},\ \bibinfo
  {pages} {485} (\bibinfo {year} {2011})}\BibitemShut {NoStop}%
\bibitem [{\citenamefont {Wang}\ \emph
  {et~al.}(2016{\natexlab{c}})\citenamefont {Wang}, \citenamefont {Hu},\ and\
  \citenamefont {Nevidomskyy}}]{Wangzhentao2016}%
  \BibitemOpen
  \bibfield  {author} {\bibinfo {author} {\bibfnamefont {Zhentao}\ \bibnamefont
  {Wang}}, \bibinfo {author} {\bibfnamefont {Wen-Jun}\ \bibnamefont {Hu}}, \
  and\ \bibinfo {author} {\bibfnamefont {Andriy~H.}\ \bibnamefont
  {Nevidomskyy}},\ }\bibfield  {title} {\enquote {\bibinfo {title} {Spin
  ferroquadrupolar order in the nematic phase of fese},}\ }\href {\doibase
  10.1103/PhysRevLett.116.247203} {\bibfield  {journal} {\bibinfo  {journal}
  {Phys. Rev. Lett.}\ }\textbf {\bibinfo {volume} {116}},\ \bibinfo {pages}
  {247203} (\bibinfo {year} {2016}{\natexlab{c}})}\BibitemShut {NoStop}%
\bibitem [{\citenamefont {Gong}\ \emph {et~al.}(2017)\citenamefont {Gong},
  \citenamefont {Zhu}, \citenamefont {Sheng},\ and\ \citenamefont
  {Yang}}]{gong2017}%
  \BibitemOpen
  \bibfield  {author} {\bibinfo {author} {\bibfnamefont {Shou-Shu}\
  \bibnamefont {Gong}}, \bibinfo {author} {\bibfnamefont {W.}~\bibnamefont
  {Zhu}}, \bibinfo {author} {\bibfnamefont {D.~N.}\ \bibnamefont {Sheng}}, \
  and\ \bibinfo {author} {\bibfnamefont {Kun}\ \bibnamefont {Yang}},\
  }\bibfield  {title} {\enquote {\bibinfo {title} {Possible nematic spin liquid
  in spin-1 antiferromagnetic system on the square lattice: Implications for
  the nematic paramagnetic state of fese},}\ }\href {\doibase
  10.1103/PhysRevB.95.205132} {\bibfield  {journal} {\bibinfo  {journal} {Phys.
  Rev. B}\ }\textbf {\bibinfo {volume} {95}},\ \bibinfo {pages} {205132}
  (\bibinfo {year} {2017})}\BibitemShut {NoStop}%
\bibitem [{\citenamefont {Lai}\ \emph {et~al.}(2017)\citenamefont {Lai},
  \citenamefont {Hu}, \citenamefont {Nica}, \citenamefont {Yu},\ and\
  \citenamefont {Si}}]{lai2017}%
  \BibitemOpen
  \bibfield  {author} {\bibinfo {author} {\bibfnamefont {Hsin-Hua}\
  \bibnamefont {Lai}}, \bibinfo {author} {\bibfnamefont {Wen-Jun}\ \bibnamefont
  {Hu}}, \bibinfo {author} {\bibfnamefont {Emilian~M.}\ \bibnamefont {Nica}},
  \bibinfo {author} {\bibfnamefont {Rong}\ \bibnamefont {Yu}}, \ and\ \bibinfo
  {author} {\bibfnamefont {Qimiao}\ \bibnamefont {Si}},\ }\bibfield  {title}
  {\enquote {\bibinfo {title} {Antiferroquadrupolar order and rotational
  symmetry breaking in a generalized bilinear-biquadratic model on a square
  lattice},}\ }\href {\doibase 10.1103/PhysRevLett.118.176401} {\bibfield
  {journal} {\bibinfo  {journal} {Phys. Rev. Lett.}\ }\textbf {\bibinfo
  {volume} {118}},\ \bibinfo {pages} {176401} (\bibinfo {year}
  {2017})}\BibitemShut {NoStop}%
\bibitem [{\citenamefont {Ruiz}\ \emph {et~al.}(2019)\citenamefont {Ruiz},
  \citenamefont {Wang}, \citenamefont {Moritz}, \citenamefont {Baum},
  \citenamefont {Hackl},\ and\ \citenamefont {Devereaux}}]{Ruiz2019}%
  \BibitemOpen
  \bibfield  {author} {\bibinfo {author} {\bibfnamefont {Harrison}\
  \bibnamefont {Ruiz}}, \bibinfo {author} {\bibfnamefont {Yao}\ \bibnamefont
  {Wang}}, \bibinfo {author} {\bibfnamefont {Brian}\ \bibnamefont {Moritz}},
  \bibinfo {author} {\bibfnamefont {Andreas}\ \bibnamefont {Baum}}, \bibinfo
  {author} {\bibfnamefont {Rudi}\ \bibnamefont {Hackl}}, \ and\ \bibinfo
  {author} {\bibfnamefont {Thomas~P.}\ \bibnamefont {Devereaux}},\ }\bibfield
  {title} {\enquote {\bibinfo {title} {Frustrated magnetism from local moments
  in fese},}\ }\href {\doibase 10.1103/PhysRevB.99.125130} {\bibfield
  {journal} {\bibinfo  {journal} {Phys. Rev. B}\ }\textbf {\bibinfo {volume}
  {99}},\ \bibinfo {pages} {125130} (\bibinfo {year} {2019})}\BibitemShut
  {NoStop}%
\bibitem [{\citenamefont {Wang}\ \emph
  {et~al.}(2016{\natexlab{d}})\citenamefont {Wang}, \citenamefont {Sun},
  \citenamefont {Cui}, \citenamefont {Song}, \citenamefont {Li}, \citenamefont
  {Yu}, \citenamefont {Lei},\ and\ \citenamefont {Yu}}]{Yuweiqiang2016}%
  \BibitemOpen
  \bibfield  {author} {\bibinfo {author} {\bibfnamefont {P.~S.}\ \bibnamefont
  {Wang}}, \bibinfo {author} {\bibfnamefont {S.~S.}\ \bibnamefont {Sun}},
  \bibinfo {author} {\bibfnamefont {Y.}~\bibnamefont {Cui}}, \bibinfo {author}
  {\bibfnamefont {W.~H.}\ \bibnamefont {Song}}, \bibinfo {author}
  {\bibfnamefont {T.~R.}\ \bibnamefont {Li}}, \bibinfo {author} {\bibfnamefont
  {Rong}\ \bibnamefont {Yu}}, \bibinfo {author} {\bibfnamefont {Hechang}\
  \bibnamefont {Lei}}, \ and\ \bibinfo {author} {\bibfnamefont {Weiqiang}\
  \bibnamefont {Yu}},\ }\bibfield  {title} {\enquote {\bibinfo {title}
  {Pressure induced stripe-order antiferromagnetism and first-order phase
  transition in fese},}\ }\href {\doibase 10.1103/PhysRevLett.117.237001}
  {\bibfield  {journal} {\bibinfo  {journal} {Phys. Rev. Lett.}\ }\textbf
  {\bibinfo {volume} {117}},\ \bibinfo {pages} {237001} (\bibinfo {year}
  {2016}{\natexlab{d}})}\BibitemShut {NoStop}%
\bibitem [{\citenamefont {Kothapalli}\ \emph {et~al.}(2016)\citenamefont
  {Kothapalli}, \citenamefont {B{\"o}hmer}, \citenamefont {Jayasekara},
  \citenamefont {Ueland}, \citenamefont {Das}, \citenamefont {Sapkota},
  \citenamefont {Taufour}, \citenamefont {Xiao}, \citenamefont {Alp},
  \citenamefont {Bud'ko}, \citenamefont {Canfield}, \citenamefont {Kreyssig},\
  and\ \citenamefont {Goldman}}]{Bohmer2016}%
  \BibitemOpen
  \bibfield  {author} {\bibinfo {author} {\bibfnamefont {K.}~\bibnamefont
  {Kothapalli}}, \bibinfo {author} {\bibfnamefont {A.~E.}\ \bibnamefont
  {B{\"o}hmer}}, \bibinfo {author} {\bibfnamefont {W.~T.}\ \bibnamefont
  {Jayasekara}}, \bibinfo {author} {\bibfnamefont {B.~G.}\ \bibnamefont
  {Ueland}}, \bibinfo {author} {\bibfnamefont {P.}~\bibnamefont {Das}},
  \bibinfo {author} {\bibfnamefont {A.}~\bibnamefont {Sapkota}}, \bibinfo
  {author} {\bibfnamefont {V.}~\bibnamefont {Taufour}}, \bibinfo {author}
  {\bibfnamefont {Y.}~\bibnamefont {Xiao}}, \bibinfo {author} {\bibfnamefont
  {E.}~\bibnamefont {Alp}}, \bibinfo {author} {\bibfnamefont {S.~L.}\
  \bibnamefont {Bud'ko}}, \bibinfo {author} {\bibfnamefont {P.~C.}\
  \bibnamefont {Canfield}}, \bibinfo {author} {\bibfnamefont {A.}~\bibnamefont
  {Kreyssig}}, \ and\ \bibinfo {author} {\bibfnamefont {A.~I.}\ \bibnamefont
  {Goldman}},\ }\bibfield  {title} {\enquote {\bibinfo {title} {Strong
  cooperative coupling of pressure-induced magnetic order and nematicity in
  fese},}\ }\href {\doibase 10.1038/ncomms12728} {\bibfield  {journal}
  {\bibinfo  {journal} {Nature Communications}\ }\textbf {\bibinfo {volume}
  {7}},\ \bibinfo {pages} {12728} (\bibinfo {year} {2016})}\BibitemShut
  {NoStop}%
\bibitem [{\citenamefont {Herbrych}\ \emph {et~al.}(2018)\citenamefont
  {Herbrych}, \citenamefont {Kaushal}, \citenamefont {Nocera}, \citenamefont
  {Alvarez}, \citenamefont {Moreo},\ and\ \citenamefont
  {Dagotto}}]{herbrych2018}%
  \BibitemOpen
  \bibfield  {author} {\bibinfo {author} {\bibfnamefont {J}~\bibnamefont
  {Herbrych}}, \bibinfo {author} {\bibfnamefont {Nitin}\ \bibnamefont
  {Kaushal}}, \bibinfo {author} {\bibfnamefont {Alberto}\ \bibnamefont
  {Nocera}}, \bibinfo {author} {\bibfnamefont {Gonzalo}\ \bibnamefont
  {Alvarez}}, \bibinfo {author} {\bibfnamefont {Adriana}\ \bibnamefont
  {Moreo}}, \ and\ \bibinfo {author} {\bibfnamefont {E}~\bibnamefont
  {Dagotto}},\ }\bibfield  {title} {\enquote {\bibinfo {title} {Spin dynamics
  of the block orbital-selective mott phase},}\ }\href
  {https://doi.org/10.1038/s41467-018-06181-6} {\bibfield  {journal} {\bibinfo
  {journal} {Nature communications}\ }\textbf {\bibinfo {volume} {9}},\
  \bibinfo {pages} {3736} (\bibinfo {year} {2018})}\BibitemShut {NoStop}%
\bibitem [{\citenamefont {Fazekas}(1999)}]{Book-Fazekas}%
  \BibitemOpen
  \bibfield  {author} {\bibinfo {author} {\bibfnamefont {P.}~\bibnamefont
  {Fazekas}},\ }\href@noop {} {\emph {\bibinfo {title} {Lecture Notes on
  Electron Correlation and Magnetism}}}\ (\bibinfo  {publisher} {World
  Scientific, Singapore},\ \bibinfo {year} {1999})\BibitemShut {NoStop}%
\bibitem [{\citenamefont {Bauer}\ \emph {et~al.}(2012)\citenamefont {Bauer},
  \citenamefont {Corboz}, \citenamefont {L\"auchli}, \citenamefont {Messio},
  \citenamefont {Penc}, \citenamefont {Troyer},\ and\ \citenamefont
  {Mila}}]{Bauer2012}%
  \BibitemOpen
  \bibfield  {author} {\bibinfo {author} {\bibfnamefont {Bela}\ \bibnamefont
  {Bauer}}, \bibinfo {author} {\bibfnamefont {Philippe}\ \bibnamefont
  {Corboz}}, \bibinfo {author} {\bibfnamefont {Andreas~M.}\ \bibnamefont
  {L\"auchli}}, \bibinfo {author} {\bibfnamefont {Laura}\ \bibnamefont
  {Messio}}, \bibinfo {author} {\bibfnamefont {Karlo}\ \bibnamefont {Penc}},
  \bibinfo {author} {\bibfnamefont {Matthias}\ \bibnamefont {Troyer}}, \ and\
  \bibinfo {author} {\bibfnamefont {Fr\'ed\'eric}\ \bibnamefont {Mila}},\
  }\bibfield  {title} {\enquote {\bibinfo {title} {Three-sublattice order in
  the su(3) heisenberg model on the square and triangular lattice},}\ }\href
  {\doibase 10.1103/PhysRevB.85.125116} {\bibfield  {journal} {\bibinfo
  {journal} {Phys. Rev. B}\ }\textbf {\bibinfo {volume} {85}},\ \bibinfo
  {pages} {125116} (\bibinfo {year} {2012})}\BibitemShut {NoStop}%
\bibitem [{\citenamefont {White}(1992)}]{White1992}%
  \BibitemOpen
  \bibfield  {author} {\bibinfo {author} {\bibfnamefont {Steven~R.}\
  \bibnamefont {White}},\ }\bibfield  {title} {\enquote {\bibinfo {title}
  {Density matrix formulation for quantum renormalization groups},}\ }\href
  {\doibase 10.1103/PhysRevLett.69.2863} {\bibfield  {journal} {\bibinfo
  {journal} {Phys. Rev. Lett.}\ }\textbf {\bibinfo {volume} {69}},\ \bibinfo
  {pages} {2863--2866} (\bibinfo {year} {1992})}\BibitemShut {NoStop}%
\bibitem [{\citenamefont {Gong}\ \emph {et~al.}(2014)\citenamefont {Gong},
  \citenamefont {Zhu}, \citenamefont {Sheng}, \citenamefont {Motrunich},\ and\
  \citenamefont {Fisher}}]{gong2014square}%
  \BibitemOpen
  \bibfield  {author} {\bibinfo {author} {\bibfnamefont {Shou-Shu}\
  \bibnamefont {Gong}}, \bibinfo {author} {\bibfnamefont {Wei}\ \bibnamefont
  {Zhu}}, \bibinfo {author} {\bibfnamefont {D.~N.}\ \bibnamefont {Sheng}},
  \bibinfo {author} {\bibfnamefont {Olexei~I.}\ \bibnamefont {Motrunich}}, \
  and\ \bibinfo {author} {\bibfnamefont {Matthew P.~A.}\ \bibnamefont
  {Fisher}},\ }\bibfield  {title} {\enquote {\bibinfo {title} {Plaquette
  ordered phase and quantum phase diagram in the spin-$\frac{1}{2}$
  ${J}_{1}\text{\ensuremath{-}}{J}_{2}$ square heisenberg model},}\ }\href
  {\doibase 10.1103/PhysRevLett.113.027201} {\bibfield  {journal} {\bibinfo
  {journal} {Phys. Rev. Lett.}\ }\textbf {\bibinfo {volume} {113}},\ \bibinfo
  {pages} {027201} (\bibinfo {year} {2014})}\BibitemShut {NoStop}%
\bibitem [{\citenamefont {Hu}\ \emph {et~al.}(2020)\citenamefont {Hu},
  \citenamefont {Gong}, \citenamefont {Lai}, \citenamefont {Si},\ and\
  \citenamefont {Dagotto}}]{hu2020}%
  \BibitemOpen
  \bibfield  {author} {\bibinfo {author} {\bibfnamefont {Wen-Jun}\ \bibnamefont
  {Hu}}, \bibinfo {author} {\bibfnamefont {Shou-Shu}\ \bibnamefont {Gong}},
  \bibinfo {author} {\bibfnamefont {Hsin-Hua}\ \bibnamefont {Lai}}, \bibinfo
  {author} {\bibfnamefont {Qimiao}\ \bibnamefont {Si}}, \ and\ \bibinfo
  {author} {\bibfnamefont {Elbio}\ \bibnamefont {Dagotto}},\ }\bibfield
  {title} {\enquote {\bibinfo {title} {Density matrix renormalization group
  study of nematicity in two dimensions: Application to a spin-1
  bilinear-biquadratic model on the square lattice},}\ }\href {\doibase
  10.1103/PhysRevB.101.014421} {\bibfield  {journal} {\bibinfo  {journal}
  {Phys. Rev. B}\ }\textbf {\bibinfo {volume} {101}},\ \bibinfo {pages}
  {014421} (\bibinfo {year} {2020})}\BibitemShut {NoStop}%
\bibitem [{\citenamefont {Ding}\ \emph {et~al.}(2019)\citenamefont {Ding},
  \citenamefont {Yu}, \citenamefont {Si},\ and\ \citenamefont
  {Abrahams}}]{ding2019}%
  \BibitemOpen
  \bibfield  {author} {\bibinfo {author} {\bibfnamefont {Wenxin}\ \bibnamefont
  {Ding}}, \bibinfo {author} {\bibfnamefont {Rong}\ \bibnamefont {Yu}},
  \bibinfo {author} {\bibfnamefont {Qimiao}\ \bibnamefont {Si}}, \ and\
  \bibinfo {author} {\bibfnamefont {Elihu}\ \bibnamefont {Abrahams}},\
  }\bibfield  {title} {\enquote {\bibinfo {title} {Effective exchange
  interactions for bad metals and implications for iron-based
  superconductors},}\ }\href {\doibase 10.1103/PhysRevB.100.235113} {\bibfield
  {journal} {\bibinfo  {journal} {Phys. Rev. B}\ }\textbf {\bibinfo {volume}
  {100}},\ \bibinfo {pages} {235113} (\bibinfo {year} {2019})}\BibitemShut
  {NoStop}%
\bibitem [{\citenamefont {Bendele}\ \emph {et~al.}(2012)\citenamefont
  {Bendele}, \citenamefont {Ichsanow}, \citenamefont {Pashkevich},
  \citenamefont {Keller}, \citenamefont {Str\"assle}, \citenamefont {Gusev},
  \citenamefont {Pomjakushina}, \citenamefont {Conder}, \citenamefont
  {Khasanov},\ and\ \citenamefont {Keller}}]{Bendele2012}%
  \BibitemOpen
  \bibfield  {author} {\bibinfo {author} {\bibfnamefont {M.}~\bibnamefont
  {Bendele}}, \bibinfo {author} {\bibfnamefont {A.}~\bibnamefont {Ichsanow}},
  \bibinfo {author} {\bibfnamefont {Yu.}\ \bibnamefont {Pashkevich}}, \bibinfo
  {author} {\bibfnamefont {L.}~\bibnamefont {Keller}}, \bibinfo {author}
  {\bibfnamefont {Th.}\ \bibnamefont {Str\"assle}}, \bibinfo {author}
  {\bibfnamefont {A.}~\bibnamefont {Gusev}}, \bibinfo {author} {\bibfnamefont
  {E.}~\bibnamefont {Pomjakushina}}, \bibinfo {author} {\bibfnamefont
  {K.}~\bibnamefont {Conder}}, \bibinfo {author} {\bibfnamefont
  {R.}~\bibnamefont {Khasanov}}, \ and\ \bibinfo {author} {\bibfnamefont
  {H.}~\bibnamefont {Keller}},\ }\bibfield  {title} {\enquote {\bibinfo {title}
  {Coexistence of superconductivity and magnetism in
  fese${}_{1\ensuremath{-}x}$ under pressure},}\ }\href {\doibase
  10.1103/PhysRevB.85.064517} {\bibfield  {journal} {\bibinfo  {journal} {Phys.
  Rev. B}\ }\textbf {\bibinfo {volume} {85}},\ \bibinfo {pages} {064517}
  (\bibinfo {year} {2012})}\BibitemShut {NoStop}%
\bibitem [{\citenamefont {Miyoshi}\ \emph {et~al.}(2014)\citenamefont
  {Miyoshi}, \citenamefont {Morishita}, \citenamefont {Mutou}, \citenamefont
  {Kondo}, \citenamefont {Seida}, \citenamefont {Fujiwara}, \citenamefont
  {Takeuchi},\ and\ \citenamefont {Nishigori}}]{Miyoshi2014}%
  \BibitemOpen
  \bibfield  {author} {\bibinfo {author} {\bibfnamefont {Kiyotaka}\
  \bibnamefont {Miyoshi}}, \bibinfo {author} {\bibfnamefont {Koh}\ \bibnamefont
  {Morishita}}, \bibinfo {author} {\bibfnamefont {Eriko}\ \bibnamefont
  {Mutou}}, \bibinfo {author} {\bibfnamefont {Masatoshi}\ \bibnamefont
  {Kondo}}, \bibinfo {author} {\bibfnamefont {Osamu}\ \bibnamefont {Seida}},
  \bibinfo {author} {\bibfnamefont {Kenji}\ \bibnamefont {Fujiwara}}, \bibinfo
  {author} {\bibfnamefont {Jun}\ \bibnamefont {Takeuchi}}, \ and\ \bibinfo
  {author} {\bibfnamefont {Shijo}\ \bibnamefont {Nishigori}},\ }\bibfield
  {title} {\enquote {\bibinfo {title} {Enhanced superconductivity on the
  tetragonal lattice in fese under hydrostatic pressure},}\ }\href {\doibase
  10.7566/JPSJ.83.013702} {\bibfield  {journal} {\bibinfo  {journal} {Journal
  of the Physical Society of Japan}\ }\textbf {\bibinfo {volume} {83}},\
  \bibinfo {pages} {013702} (\bibinfo {year} {2014})}\BibitemShut {NoStop}%
\bibitem [{\citenamefont {Kaluarachchi}\ \emph {et~al.}(2016)\citenamefont
  {Kaluarachchi}, \citenamefont {Taufour}, \citenamefont {B\"ohmer},
  \citenamefont {Tanatar}, \citenamefont {Bud'ko}, \citenamefont {Kogan},
  \citenamefont {Prozorov},\ and\ \citenamefont {Canfield}}]{Kaluarachchi2016}%
  \BibitemOpen
  \bibfield  {author} {\bibinfo {author} {\bibfnamefont {Udhara~S.}\
  \bibnamefont {Kaluarachchi}}, \bibinfo {author} {\bibfnamefont {Valentin}\
  \bibnamefont {Taufour}}, \bibinfo {author} {\bibfnamefont {Anna~E.}\
  \bibnamefont {B\"ohmer}}, \bibinfo {author} {\bibfnamefont {Makariy~A.}\
  \bibnamefont {Tanatar}}, \bibinfo {author} {\bibfnamefont {Sergey~L.}\
  \bibnamefont {Bud'ko}}, \bibinfo {author} {\bibfnamefont {Vladimir~G.}\
  \bibnamefont {Kogan}}, \bibinfo {author} {\bibfnamefont {Ruslan}\
  \bibnamefont {Prozorov}}, \ and\ \bibinfo {author} {\bibfnamefont {Paul~C.}\
  \bibnamefont {Canfield}},\ }\bibfield  {title} {\enquote {\bibinfo {title}
  {Nonmonotonic pressure evolution of the upper critical field in
  superconducting fese},}\ }\href {\doibase 10.1103/PhysRevB.93.064503}
  {\bibfield  {journal} {\bibinfo  {journal} {Phys. Rev. B}\ }\textbf {\bibinfo
  {volume} {93}},\ \bibinfo {pages} {064503} (\bibinfo {year}
  {2016})}\BibitemShut {NoStop}%
\bibitem [{\citenamefont {Sun}\ \emph {et~al.}(2016)\citenamefont {Sun},
  \citenamefont {Matsuura}, \citenamefont {Ye}, \citenamefont {Mizukami},
  \citenamefont {Shimozawa}, \citenamefont {Matsubayashi}, \citenamefont
  {Yamashita}, \citenamefont {Watashige}, \citenamefont {Kasahara},
  \citenamefont {Matsuda}, \citenamefont {Yan}, \citenamefont {Sales},
  \citenamefont {Uwatoko}, \citenamefont {Cheng},\ and\ \citenamefont
  {Shibauchi}}]{Sun2015}%
  \BibitemOpen
  \bibfield  {author} {\bibinfo {author} {\bibfnamefont {J.~P.}\ \bibnamefont
  {Sun}}, \bibinfo {author} {\bibfnamefont {K.}~\bibnamefont {Matsuura}},
  \bibinfo {author} {\bibfnamefont {G.~Z.}\ \bibnamefont {Ye}}, \bibinfo
  {author} {\bibfnamefont {Y.}~\bibnamefont {Mizukami}}, \bibinfo {author}
  {\bibfnamefont {M.}~\bibnamefont {Shimozawa}}, \bibinfo {author}
  {\bibfnamefont {K.}~\bibnamefont {Matsubayashi}}, \bibinfo {author}
  {\bibfnamefont {M.}~\bibnamefont {Yamashita}}, \bibinfo {author}
  {\bibfnamefont {T.}~\bibnamefont {Watashige}}, \bibinfo {author}
  {\bibfnamefont {S.}~\bibnamefont {Kasahara}}, \bibinfo {author}
  {\bibfnamefont {Y.}~\bibnamefont {Matsuda}}, \bibinfo {author} {\bibfnamefont
  {J.~Q.}\ \bibnamefont {Yan}}, \bibinfo {author} {\bibfnamefont {B.~C.}\
  \bibnamefont {Sales}}, \bibinfo {author} {\bibfnamefont {Y.}~\bibnamefont
  {Uwatoko}}, \bibinfo {author} {\bibfnamefont {J.~G.}\ \bibnamefont {Cheng}},
  \ and\ \bibinfo {author} {\bibfnamefont {T.}~\bibnamefont {Shibauchi}},\
  }\bibfield  {title} {\enquote {\bibinfo {title} {Dome-shaped magnetic order
  competing with high-temperature superconductivity at high pressures in
  fese},}\ }\href {\doibase 10.1038/ncomms12146} {\bibfield  {journal}
  {\bibinfo  {journal} {Nature Communications}\ }\textbf {\bibinfo {volume}
  {7}},\ \bibinfo {pages} {12146} (\bibinfo {year} {2016})}\BibitemShut
  {NoStop}%
\bibitem [{\citenamefont {Terashima}\ \emph {et~al.}(2016)\citenamefont
  {Terashima}, \citenamefont {Kikugawa}, \citenamefont {Kiswandhi},
  \citenamefont {Graf}, \citenamefont {Choi}, \citenamefont {Brooks},
  \citenamefont {Kasahara}, \citenamefont {Watashige}, \citenamefont {Matsuda},
  \citenamefont {Shibauchi}, \citenamefont {Wolf}, \citenamefont {B\"ohmer},
  \citenamefont {Hardy}, \citenamefont {Meingast}, \citenamefont
  {L\"ohneysen},\ and\ \citenamefont {Uji}}]{Terashima2016}%
  \BibitemOpen
  \bibfield  {author} {\bibinfo {author} {\bibfnamefont {Taichi}\ \bibnamefont
  {Terashima}}, \bibinfo {author} {\bibfnamefont {Naoki}\ \bibnamefont
  {Kikugawa}}, \bibinfo {author} {\bibfnamefont {Andhika}\ \bibnamefont
  {Kiswandhi}}, \bibinfo {author} {\bibfnamefont {David}\ \bibnamefont {Graf}},
  \bibinfo {author} {\bibfnamefont {Eun-Sang}\ \bibnamefont {Choi}}, \bibinfo
  {author} {\bibfnamefont {James~S.}\ \bibnamefont {Brooks}}, \bibinfo {author}
  {\bibfnamefont {Shigeru}\ \bibnamefont {Kasahara}}, \bibinfo {author}
  {\bibfnamefont {Tatsuya}\ \bibnamefont {Watashige}}, \bibinfo {author}
  {\bibfnamefont {Yuji}\ \bibnamefont {Matsuda}}, \bibinfo {author}
  {\bibfnamefont {Takasada}\ \bibnamefont {Shibauchi}}, \bibinfo {author}
  {\bibfnamefont {Thomas}\ \bibnamefont {Wolf}}, \bibinfo {author}
  {\bibfnamefont {Anna~E.}\ \bibnamefont {B\"ohmer}}, \bibinfo {author}
  {\bibfnamefont {Fr\'ed\'eric}\ \bibnamefont {Hardy}}, \bibinfo {author}
  {\bibfnamefont {Christoph}\ \bibnamefont {Meingast}}, \bibinfo {author}
  {\bibfnamefont {Hilbert~v.}\ \bibnamefont {L\"ohneysen}}, \ and\ \bibinfo
  {author} {\bibfnamefont {Shinya}\ \bibnamefont {Uji}},\ }\bibfield  {title}
  {\enquote {\bibinfo {title} {Fermi surface reconstruction in fese under high
  pressure},}\ }\href {\doibase 10.1103/PhysRevB.93.094505} {\bibfield
  {journal} {\bibinfo  {journal} {Phys. Rev. B}\ }\textbf {\bibinfo {volume}
  {93}},\ \bibinfo {pages} {094505} (\bibinfo {year} {2016})}\BibitemShut
  {NoStop}%
\end{thebibliography}%

\end{document}